\documentclass[10pt,conference]{IEEEtran}
\IEEEoverridecommandlockouts

\usepackage{cite}
\usepackage{amsmath,amssymb,amsfonts}
\usepackage{algorithmic}
\usepackage{graphicx}
\usepackage{textcomp}
\usepackage{booktabs}
\usepackage{xcolor}
\usepackage{pifont}
\usepackage{tcolorbox}
\usepackage{xurl}
\usepackage{lineno}
\usepackage{pifont}
\newcommand{\benchmark}{Interaction2Code}
\newcommand{\task}{Interaction-to-Code}

\usepackage{enumerate}

\usepackage{adjustbox}

\definecolor{deepgreen}{rgb}{0, 0.6, 0.2}
\definecolor{deeepyellow}{rgb}{1, 0.85, 0.4}

\newcommand{\redcross}{\begin{tikzpicture}
  \fill[red] (0,0) circle (0.12cm);
\end{tikzpicture}}

\newcommand{\partialcheck}{\begin{tikzpicture}
  \fill[deeepyellow] (0,0) circle (0.12cm);
\end{tikzpicture}}
\usepackage{xcolor}

\usepackage{times}
\usepackage{latexsym}
\usepackage{booktabs} 

\usepackage{listings}
\usepackage{xcolor}
\usepackage{subfigure}
\usepackage{multirow}
\usepackage{colortbl}
\usepackage{pifont}
\usepackage{tikz}
\usepackage{tcolorbox}
\usepackage{enumitem}

\usepackage{algorithm}
\usepackage{algorithmic}

\usepackage[T1]{fontenc}

\usepackage[utf8]{inputenc}

\usepackage{microtype}

\usepackage{inconsolata}

\usepackage{graphicx}

\newcommand{\yes}{\color{green!60!black}\ding{51}}
\newcommand{\no}{\color{red!60!black}\ding{55}} 
\usepackage{titletoc}

\newtcolorbox{myquote}{
    width=0.48\textwidth,
    colback=white,
    colframe=black,
    fontupper=\itshape,
    boxrule=0.2mm,
    left=1mm,
    right=1mm,
    arc=3mm,
    auto outer arc
}

\def\BibTeX{{\rm B\kern-.05em{\sc i\kern-.025em b}\kern-.08em
    T\kern-.1667em\lower.7ex\hbox{E}\kern-.125emX}}
\begin{document}


\title{Interaction2Code: Benchmarking MLLM-based Interactive Webpage Code Generation from Interactive Prototyping}


\author{\IEEEauthorblockN{Jingyu Xiao$^{1}$, Yuxuan Wan$^{1}$, Yintong Huo$^{2\ast}$, Zixin Wang$^{1}$, Xinyi Xu$^{1}$, Wenxuan Wang$^{3}$,\\ Zhiyao Xu$^{4}$, Yuhang Wang$^{5}$, Michael R. Lyu$^{1}$}

\IEEEauthorblockA{$^1$The Chinese University of Hong Kong, China $^2$ Singapore Management University, Singapore}

\IEEEauthorblockA{$^3$ Renmin University of China, China $^4$ Tsinghua University, China $^5$ Southwest University, China}

\IEEEauthorblockA{\{jyxiao, yxwan\}@link.cuhk.edu.hk, ythuo@smu.edu.sg, \{zixinwang, xinyixu\}@link.cuhk.edu.hk, jwxwang@gmail.com, \\ xu-zy25@mails.tsinghua.edu.cn, wyh20030323@email.swu.edu.cn, lyu@cse.cuhk.edu.hk}
\thanks{$^{\ast}$ Yintong Huo is the corresponding author.}
}



\maketitle

\begin{abstract}

    Multimodal Large Language Models (MLLMs) have demonstrated remarkable performance on the design-to-code task, i.e., generating UI code from UI mock-ups. However, existing benchmarks only contain static web pages for evaluation and ignore the dynamic interaction, limiting the practicality, usability and user engagement of the generated webpages.

    To bridge these gaps, we present the first systematic investigation of MLLMs in generating interactive webpages. Specifically, we formulate the \textbf{\task} task and establish the \textbf{\benchmark} benchmark, encompassing \textbf{127} unique webpages and \textbf{374} distinct interactions across \textbf{15} webpage types and \textbf{31} interaction categories. Through comprehensive experiments utilizing state-of-the-art (SOTA) MLLMs, evaluated via both automatic metrics and human assessments, we identify four critical limitations of MLLM on \task \ task: (1) inadequate generation of interaction compared with full page, (2) prone to ten types of failure, (3) poor performance on visually subtle interactions, and (4) insufficient undestanding on interaction when limited to single-modality visual descriptions. To address these limitations, we propose four enhancement strategies: \textbf{Interactive Element Highlighting}, \textbf{Failure-aware Prompting (FAP)}, \textbf{Visual Saliency Enhancement}, and \textbf{Visual-Textual Descriptions Combination}, all aiming at improving MLLMs' performance on the \task \ task. Our data and code are available in \url{https://github.com/WebPAI/Interaction2Code}.

\end{abstract}


\section{Introduction}

As of 2025, the digital landscape consists of approximately 1.09 billion websites \cite{statistics2024}, supporting a wide range of applications in daily life. The design and development of Graphical User Interfaces (GUIs) play a critical role in shaping a website’s aesthetics and functionality, improving user satisfaction through effective layout, colors, and typography \cite{Chen2018FromUI, Kuusinen2013DesigningUE}. However, translating these designs into functional code is complex and time-consuming, often leading to discrepancies between the design and final implementation \cite{Chen2018FromUI, Nguyen2015ReverseEM, Lelli2015ClassifyingAQ, Moran2018AutomatedRO, Zeidler2013EvaluatingDM}. To address this, automated GUI code generation methods have been proposed, categorized into learning-based and LLM-based approaches. Learning-based methods like Pix2code \cite{beltramelli2018pix2code} use CNNs and LSTMs to reverse-engineer GUI code from images, while Chen et al. \cite{chen2018ui} present a neural machine translator for encoding UI features and generating GUI skeletons. However, these methods struggle with generalizing to diverse web elements. Recent advances in Multimodal Large Language Models (MLLMs), such as GPT-4o \cite{openai_gpt4o}, Claude-3.5 \cite{anthropic2023}, and Gemini-1.5 \cite{google_gemini_api}, have improved visual understanding tasks \cite{Yang2023TheDO, Dai2023InstructBLIPTG, liu2025WebRSSBench} and shown remarkable performance in code generation \cite{Yu2023CoderEvalAB, Li2023EnablingPT, Du2023ClassEvalAM, lu2025exploring, lu2025next, tang2025slidecoder}. These advances open new possibilities for the Design-to-Code task, where MLLMs generate code from screenshots to replicate web elements, layout, text, and colors. For instance, Design2Code \cite{si2024design2code} uses prompts to guide MLLMs in web content understanding and code generation, while DCGen \cite{wan2024automatically} proposes a divide-and-conquer approach to improve the accuracy of generated webpage elements.

\begin{figure}[ht]
    \subfigure[Interactive elements.]{
    \label{fig:inter_ex}
    \centering
    \includegraphics[width = .19\textwidth]{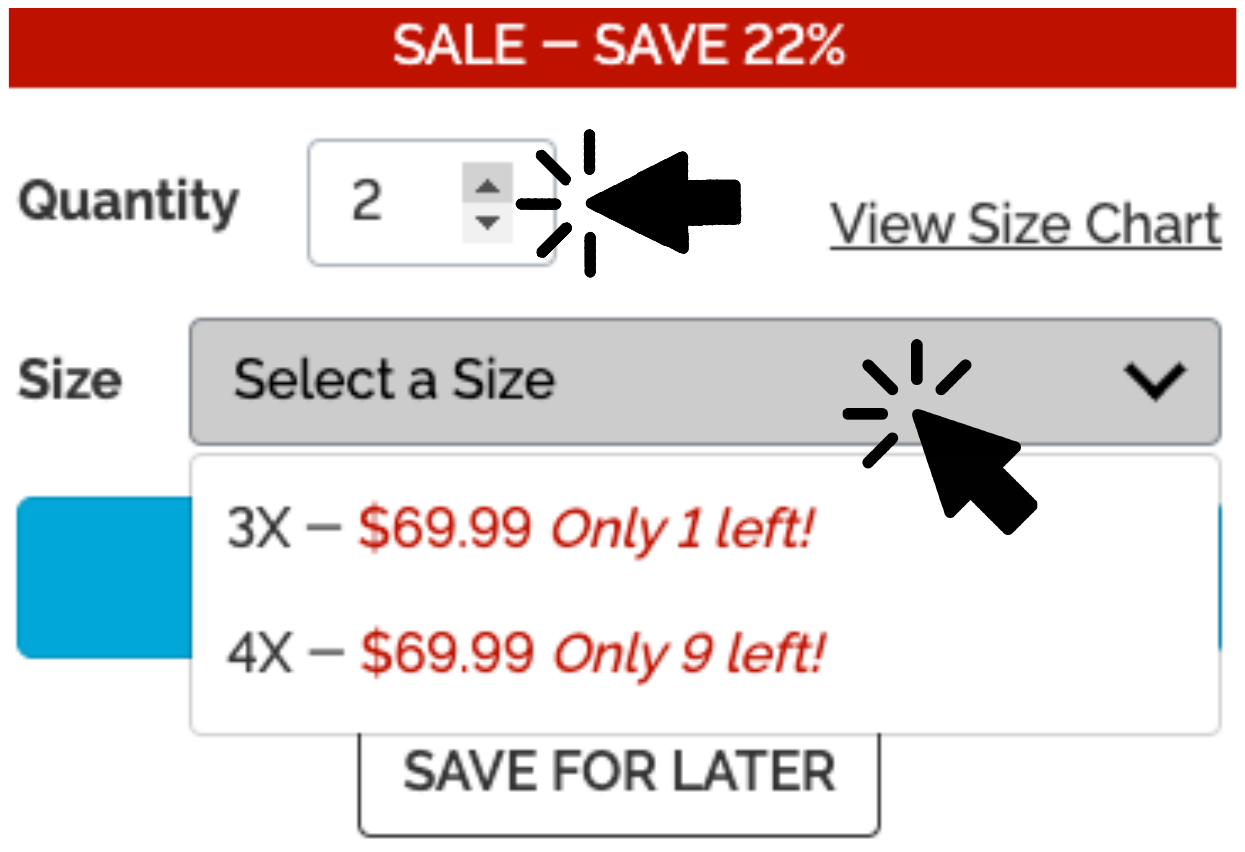}
    }
    \subfigure[Interactive and static ratio.]{
    \label{fig:ratio}
    \centering
    \includegraphics[width = .275\textwidth]{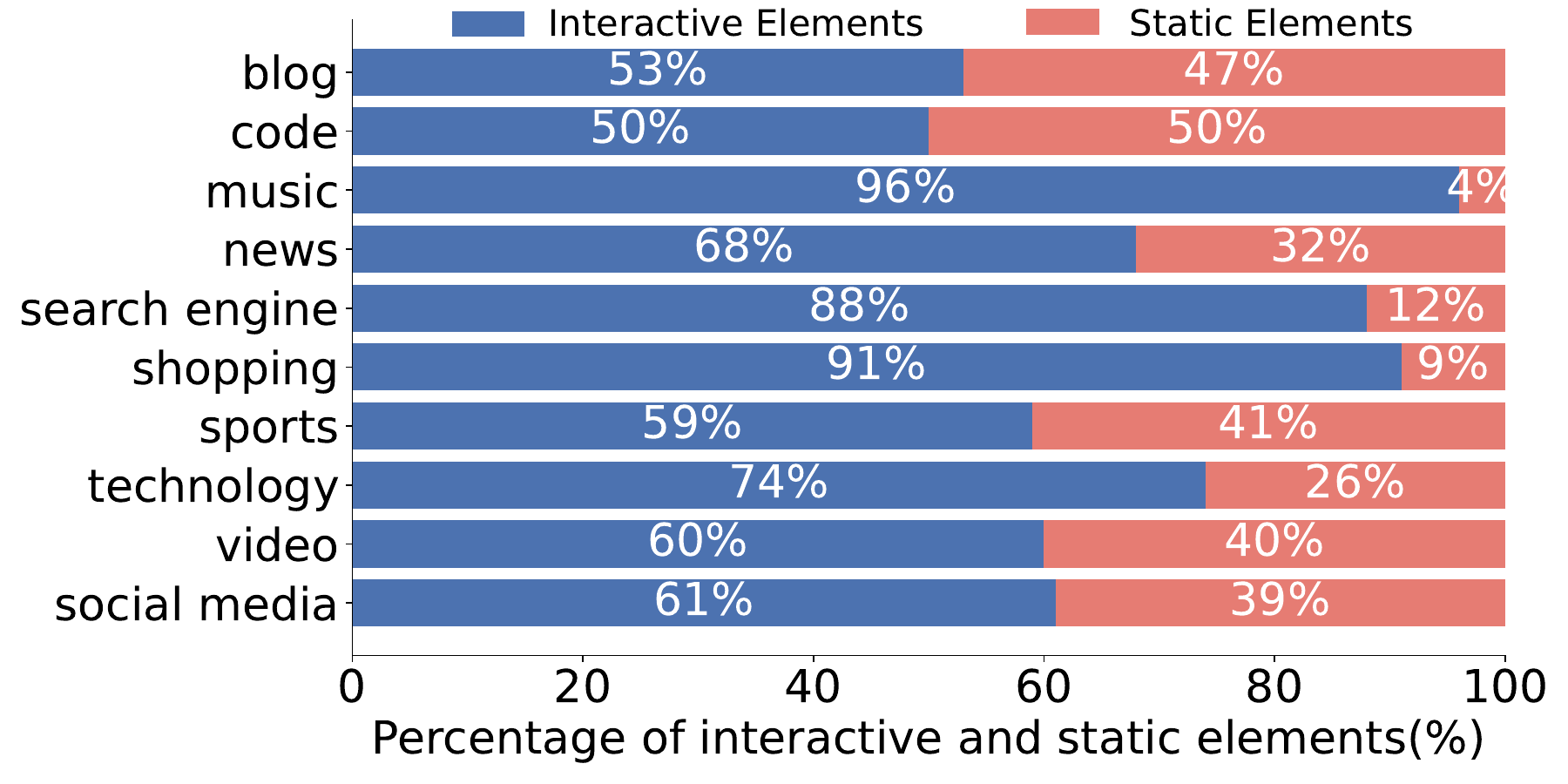}
    }
    \caption{Interaction example and interactive elements ratio of different types of webpages.}
\end{figure}

However, existing research only focuses on the static appearance of a webpage (e.g., color, layouts)\cite{si2024design2code, yun2024web2code, gui2024vision2ui, xiao2025designbench, wan2024mrweb}, ignoring the dynamic interactive properties and functionality of such elements, like size selection list, and quantity adjustment button shown in Figure~\ref{fig:inter_ex}. Additionally, we observe that such interactive elements account for a large proportion of the webpage in real-world software practices. We randomly select 10 real-world webpages with different topics to analyze the ratio of interactive elements, the results in Figure~\ref{fig:ratio} indicate that interactive elements take up more than 50\% cases.

Static webpages limit user interaction with web elements, hindering access to new content (such as browsing images via carousel buttons) or impeding task completion (like selecting clothing sizes from drop-down menus), thereby impairing overall user experience. Therefore, we argue that \textbf{a benchmark for interactive webpages is essential to enhance the practicality, usability, and user engagement of studies on auto-generated GUI code.} To this end, we provide the first systematic analysis of MLLMs' capability in generating interactive webpages. Our contributions are summarized as follows:

\begin{itemize}[leftmargin=*]
    \item \textbf{Task formulation}. We are the \textit{first} to formulate the \textbf{\task} task and present a systematic study on the code generation capabilities of MLLMs for dynamic interaction of webpages.
    \item \textbf{Benchmark}. We build the \textit{first} \textbf{real-world} webpage interaction datasets \textbf{\benchmark} \ containing \textbf{127} webpages and \textbf{374} interactions, spanning \textbf{15} webpage topics and \textbf{31} interaction types. We also provide \textbf{failure annotations} for the MLLM-generated webpages.
    



    \item  \textbf{Key Findings.} Our in-depth analysis reveals four main limitations: (1) MLLMs struggle to generate interactive part compared with full static webpage generation; (2) MLLMs are prone to make 10 types of failures; (3) MLLMs perform poorly on interactions that are not visually obvious; (4) Single visual modality description is not enough for MLLMs to understand the interaction.
    


    

    \item \textbf{Improvements.} We propose four methods to improve the performance of MLLMs on the \task \ task. (1) \textbf{Interactive element highlighting}, i.e., applying visual markers for interactive elements can improve MLLMs' performance by forcing MLLMs to focus on the Interaction. (2) \textbf{Failure-aware prompting (FAP)} can make MLLMs avoid potential failures by incorporating the failure example into prompts. (3) \textbf{Visual saliency enhancement (VSE)} allows the model to better perceive the interaction area, thereby improving the performance of interaction generation. (4) \textbf{Visual and textual description combination} can makes MLLMs understand the interaction better.

    

    

\end{itemize}




\section{Background}

\subsection{Related Work}


UI code generation techniques fall into three categories: Deep Learning (DL)-based, Computer Vision (CV)-based, and Multimodal Large Language Model (MLLM)-based methods. (1) DL-based methods: \cite{acsirouglu2019automatic, cizotto2023web, moran2018machine, Xu2021Image2e, Chen2018FromUI} use CNNs to prototype GUIs automatically. Pix2code \cite{beltramelli2018pix2code} combines CNNs and LSTM to generate a domain-specific language (DSL) from GUI images. \cite{chen2022code} uses an encoder-decoder framework with attention mechanisms to generate DSL. (2) CV-based methods: Sketch2Code \cite{jain2019sketch2code} feeds hand-drawn sketches into object detection models to generate code. REMAUI \cite{nguyen2015reverse} uses OCR to identify UI elements and generate corresponding source code. (3) MLLM-based methods: Design2Code \cite{si2024design2code} leverages MLLMs for UI code generation using text-augmented and self-revision prompting. To address issues like element omission and misarrangement, DCGen \cite{wan2024automatically} adopts a divide-and-conquer approach, generating submodules separately before assembling them. DeclarUI \cite{zhou2024bridging} uses element segmentation and page transition graphs to guide MLLMs in generating mobile app UI with jump logic. While these approaches show strong performance, none focus on generating interactive elements. EfficientUICoder~\cite{xiao2025efficientuicoder} applies token compression method to accelerate UI code generation. \textbf{Although the above works achieve decent performance on the UI2Code task, none of them consider the generation of interactive webpages.}





\begin{table}[t]
\centering
\caption{Comparisons between \benchmark \ and existing UI2Code benchmarks.}
\label{table:comparison}
\resizebox{\linewidth}{!}{
\setlength{\tabcolsep}{.12em}{
\begin{tabular}{lcccc}
\toprule
Benchmark & \begin{tabular}[c]{@{}c@{}}Real\\ World\end{tabular} & \begin{tabular}[c]{@{}c@{}}Failure\\ Annotation\end{tabular} & Interactive \\
\midrule
WebSight \cite{laurençon2024unlocking} & \no & \no & \no \\
Vision2UI \cite{gui2024vision2ui} & \yes & \no & \no \\
Design2Code \cite{si2024design2code} & \yes & \no & \no \\
DesignBench \cite{xiao2025designbench} & \yes & \no & \no \\
\textbf{Interaction2Code (Ours)} & \yes & \yes & \yes \\
\bottomrule
\end{tabular}}}
\end{table}



\subsection{Problem Definition}

\begin{figure*}[t]
    \centering
    \begin{minipage}[t]{0.185\textwidth}
        \centering
        \includegraphics[width=\linewidth]{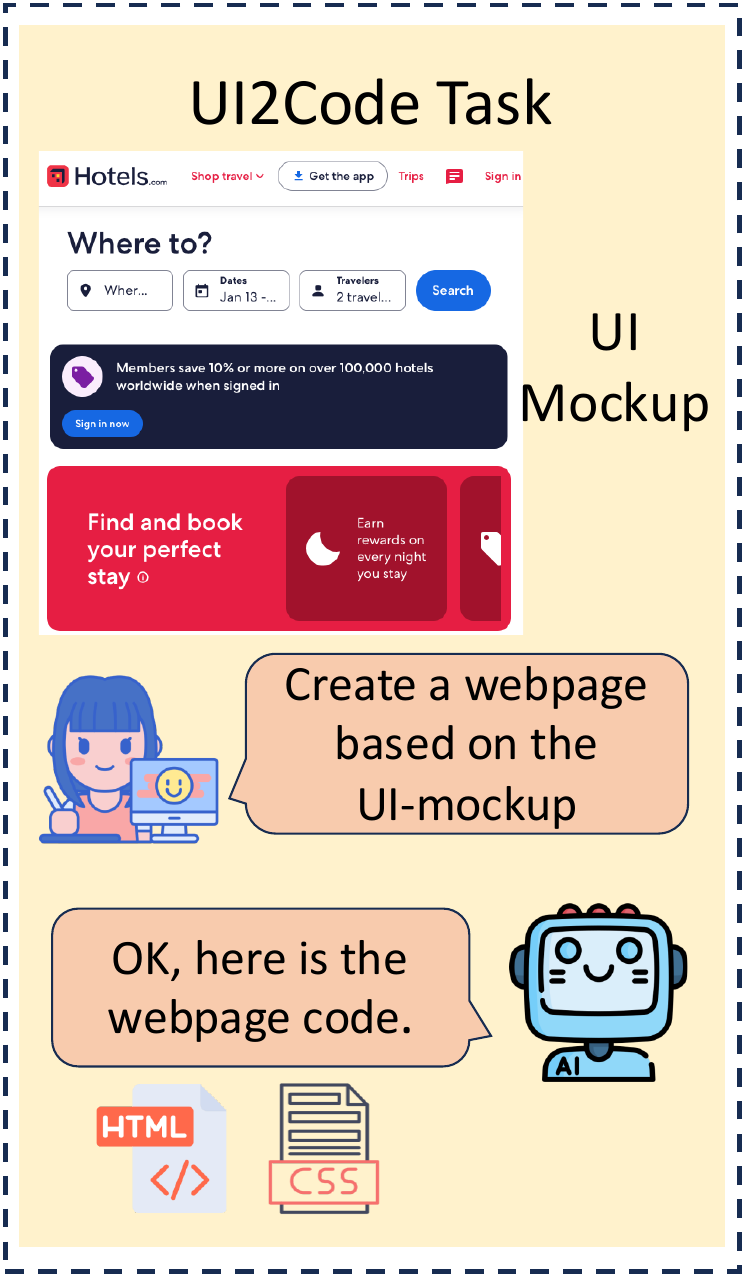}
        \caption{UI2Code.}
        \label{fig:ui2code}
    \end{minipage}
    \begin{minipage}[t]{0.79\textwidth}
        \centering
        \includegraphics[width=\linewidth]{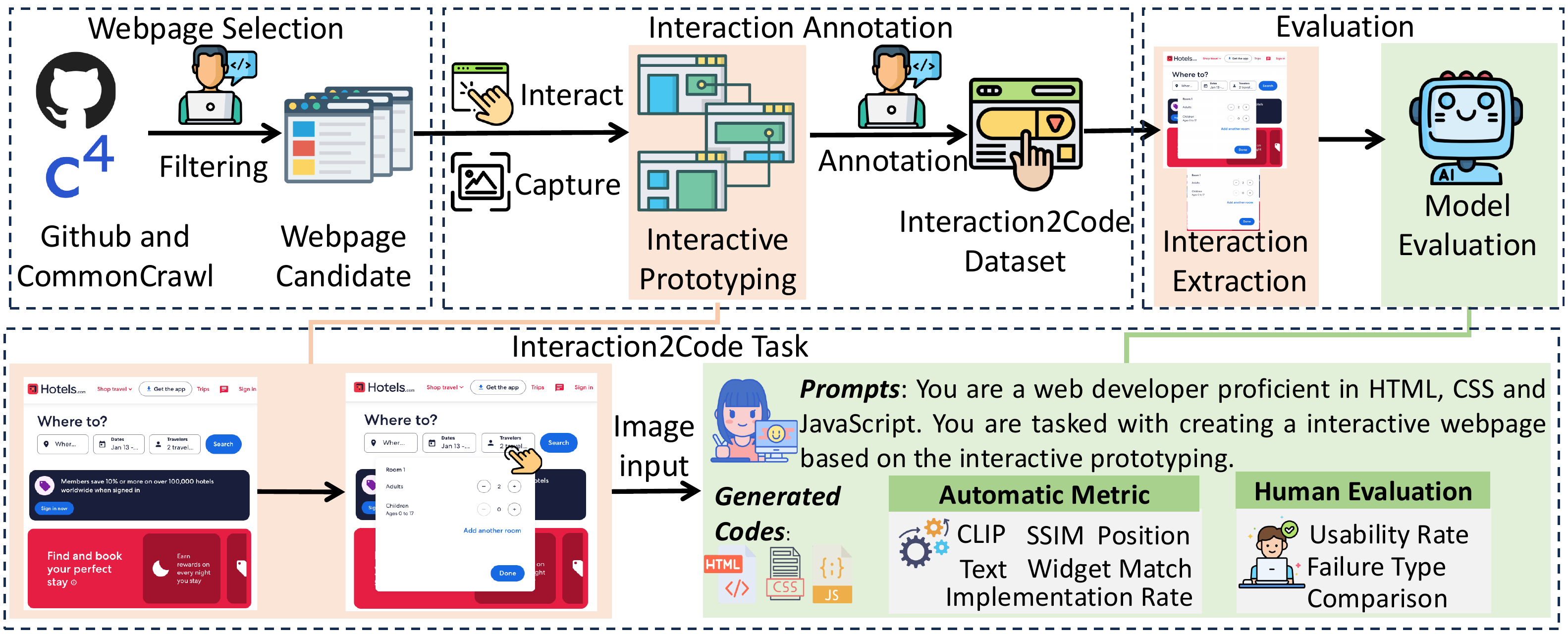}
        \caption{The construction of \benchmark \ benchmark.}
        \label{fig:overview}
    \end{minipage}
\end{figure*}

\textbf{UI-Mockup}~\cite{uimockup} is a visual representation of a user interface, essentially a static image showing the look and feel of a webpage.  Figure~\ref{fig:ui2code} shows that the UI2Code task takes the static UI-Mockup $S$ as input and generates a static webpage. \textbf{Interactive Prototyping}~\cite{interactiveprototyping} is a functional model of that design, allowing users to simulate interactions and navigate through the interface to test usability and functionality before full development. An interactive behavior is represented as an interactive prototype $IP=\{S_o, S_I\}$, where $S_o$ is the UI-Mockup of original webpage and $S_I$ is the UI-Mockup after the interaction $I$.

\textbf{\benchmark \ task} takes the interactive prototyping $IP$ as input and generates an interactive webpage as shown in Figure~\ref{fig:overview}.

\section{The \benchmark \ Benchmark}




\subsection{Dataset Collection}



We follow these steps for constructing benchmark that represent a variety of real-world use cases (i.e., diverse webpages and interactions).



\textbf{Webpage Selection.} We collect webpages from CommonCrawl (C4 validation set \cite{raffel2020exploring}) and GitHub. (1) CommonCrawl: Following the Design2Code approach \cite{si2024design2code}, we automatically filter out webpages that are either too long or too simple (containing only images or text) and perform deduplication. We then select 15 common web topics and randomly sample 1,000 webpages related to these topics. Four PhD students majoring in computer science, each with front-end development experience, are tasked with selecting approximately 25 webpages each, resulting in 100 webpages from C4. The selection criteria are: 1) complexity: each webpage must include at least one meaningful interaction; 2) diversity: the selection aims to cover a broad range of webpages with different interaction types. Since most C4 websites are traditional and do not use UI frameworks, we also collect webpages from GitHub projects that use UI frameworks. (2) GitHub projects: We search for ``open-production-web-projects'' and ``awesome-opensource-apps'' on GitHub to compile a list of web projects, from which we identify 27 popular projects with deployed links and high star counts. These projects, averaging 13k stars, represent various real-world website uses, ranging from commercial product frontends to blogs, demonstrating their quality and popularity. Ultimately, we compile a dataset of 127 webpages. Detailed selection guidelines are provided in our artifact.

\textbf{Interaction Annotation.} (1) Interactive Prototyping Construction. In real-world webpages, there are many trivial interactions, like underlining texts when hovering. To preserve meaningful interactions and ensure the complexity and diversity of interactions, the four PhD students are employed to interact with webpages and select complex and meaningful interactions to capture the pre- and post-interaction screenshots to build interactive prototyping (the guideline is shown in our artifact). Finally, 1-10 important and functional interactions are retained on one webpage and we get 374 interactions. (2) Annotation. The four PhD students manually annotate the topics of the web pages, the development framework, and the types of interactions for benchmark diversity analysis.




\subsection{Data Statistics and Diversity}

\begin{figure}[t]
    \subfigure[Topic.]{
    \label{fig:domain}
    \centering
    \includegraphics[width = .25\textwidth]{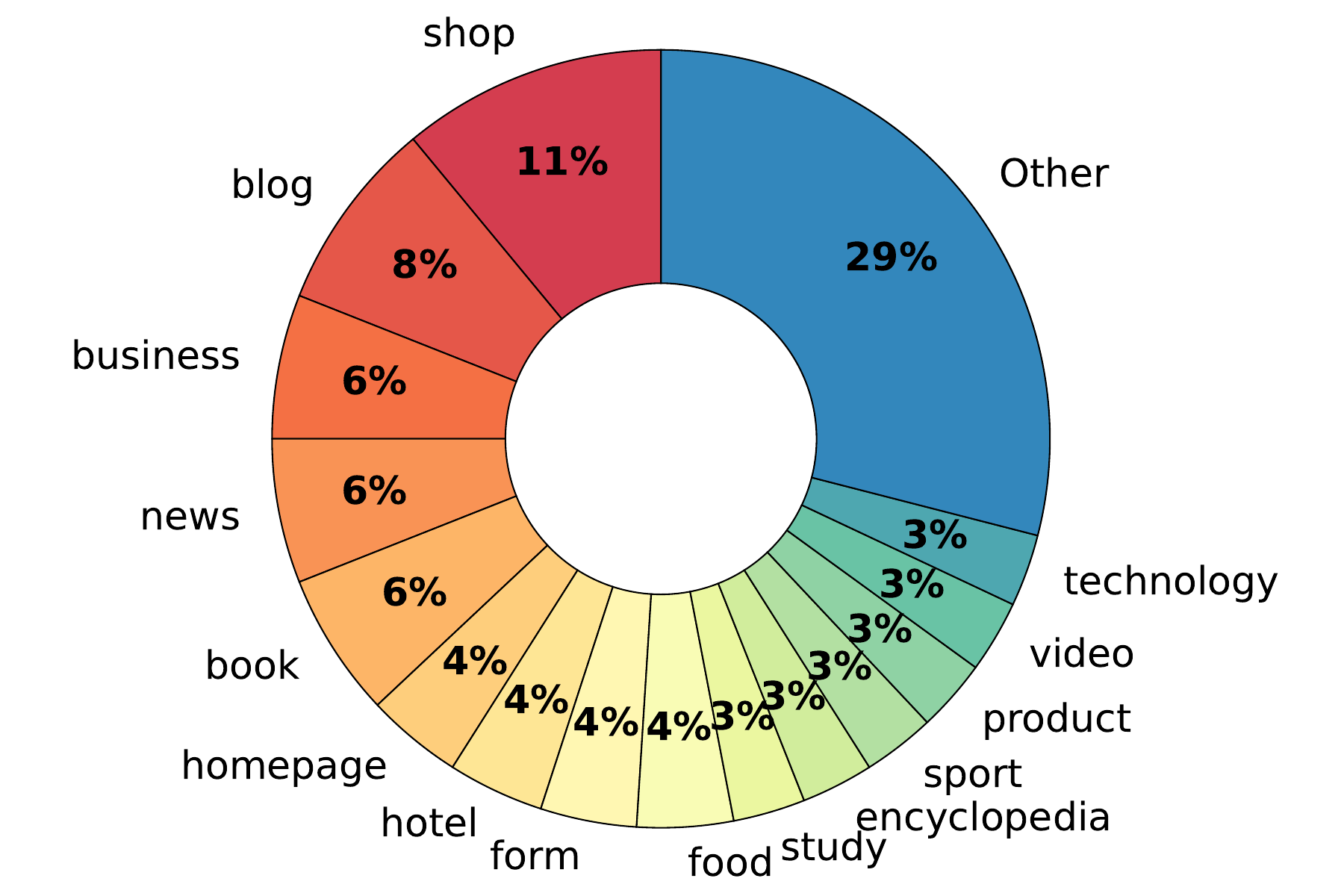}
    }
    \subfigure[Framework.]{
    \label{fig:frame}
    \centering
    \includegraphics[width = .19\textwidth]{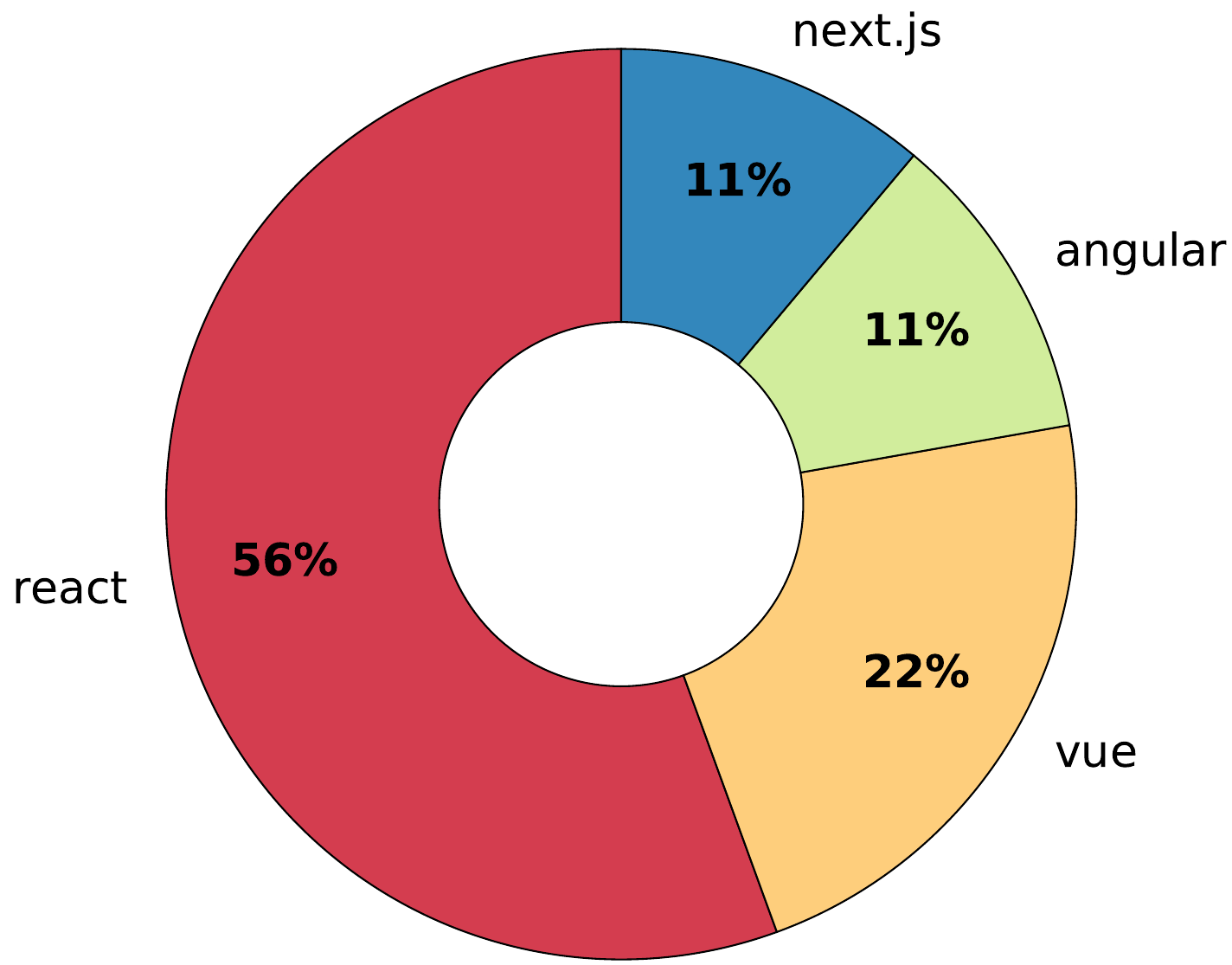}
    }
    \caption{Topic and framework distribution.}
\end{figure}

\textbf{Topic and Framework Distribution.} Figure~\ref{fig:domain} shows that our benchmark covers a diverse range of web topics with more than 15 types, including business, shop, technology, entertainment, and so on. Figure~\ref{fig:frame} shows that the benchmark includes mainstream front-end open source frameworks such as react, next.js, vue, and angular.


\begin{table}[ht]
\centering
\small
\setlength{\tabcolsep}{0.12em}
\caption{Tag and visual categories distribution.}
\label{tab:interact_combined}
\begin{tabular}{@{}lc|lc|lc@{}}
\toprule
\multicolumn{4}{c|}{Tag Categories}  & \multicolumn{2}{c}{Visual Categories} \\   \midrule
Element & Number & Element & Number & Type & Number \\
\midrule
button & 235 & summary & 15 & text & 162 \\
input & 52 & form & 13 & new component & 161 \\
span & 37 & detail & 12 & color & 85 \\
link & 36 & video & 11 & position & 45 \\
select & 35 & area & 9 & switch & 41 \\
textarea & 35 & output & 9 & new page & 37 \\
option & 31 & datalist & 8 & new window & 34 \\
iframe & 28 & dialog & 6 & size & 20 \\
text & 24 & audio & 5 & - & - \\
progress & 22 & template & 3 & - & - \\
image & 21 & table & 1 & - & - \\
label & 16 & - & - & - & - \\
\bottomrule
\end{tabular}
\end{table}

\textbf{Interaction Type Distribution.} We manually annotate the type of interactions based on their element tag and the visual effect perspective. Tag categories come from HTML tags such as button, image, and link. Buttons, input boxes, and links are the most frequent types as shown in Table~\ref{tab:interact_combined} and play a great role in human-website interaction. Visual categories involve changes in color, size, position, text, etc. Note that one interaction may belong to multiple tag categories and visual categories. Table~\ref{tab:interact_combined} demonstrates that \benchmark \ benchmark has a rich set of interaction types, including  23 tag categories and 8 visual categories.

\subsection{Evaluation}
\label{subsec:evaluation}

\textbf{Automated Interaction} When generating web pages, we prompt MLLMs to encode the id of the interactive elements (for example, id="interact1"). During evaluation, we apply selenium webdriver \cite{selenium} to locate the interactive elements by id and automatically interact with the generated webpage and take screenshots to construct the interactive prototyping.

\textbf{Interaction Extraction.} After obtaining the interactive prototyping (i.e., screenshots before and after the interaction), we automatically extract the interactive part for evaluation. For interactions that preserve webpage dimensions, we identify affected areas through pixel-wise subtraction. For interactions that alter webpage dimensions (e.g., showing details), we employ the Git diff tool~\cite{gitdiff} to locate modified rows and columns, with their intersections marking the affected regions. The algorithm is shown in Algorithm~\ref{alg:region_detection}.

\begin{algorithm}
\footnotesize
\caption{Interaction Part Extraction Algorithm}
\label{alg:region_detection}
\begin{algorithmic}[1]
\REQUIRE
    \STATE Webpage screenshot $A$ (Before interaction)
    \STATE Webpage screenshot $B$ (After interaction)
\ENSURE
    \STATE Coordinates $(x_{min}, y_{min}, x_{max}, y_{max})$ of interaction region
    \IF{$dim(A) = dim(B)$}
        \STATE $D \gets |A - B|$;
        \STATE $C \gets \{(x,y) | D(x,y) \neq 0\}$;
        \STATE $x_{min} \gets \min\{x | (x,y) \in C\}$;
        \STATE $x_{max} \gets \max\{x | (x,y) \in C\}$;
        \STATE $y_{min} \gets \min\{y | (x,y) \in C\}$;
        \STATE $y_{max} \gets \max\{y | (x,y) \in C\}$;
    \ELSE
        \STATE $diff\_rows \gets$ DiffTool($A, B$);
        \STATE $diff\_cols \gets$ DiffTool($A^{T}, B^{T}$);
        \STATE $x_{min} \gets \min(diff\_cols)$;
        \STATE $x_{max} \gets \max(diff\_cols)$;
        \STATE $y_{min} \gets \min(diff\_rows)$;
        \STATE $y_{max} \gets \max(diff\_rows)$;
    \ENDIF
    \STATE \RETURN $(x_{min}, y_{min}, x_{max}, y_{max})$
\end{algorithmic}
\end{algorithm}
\textbf{Full Page Metrics}. We measure the quality of generated webpages from the following perspectives: \textbf{(1) Visual Similarity}. We use CLIP score~\cite{radford2021learning} to measure the visual similarity. \textbf{(2) Structure Similarity}. SSIM \cite{wang2004image} (Structural Similarity Index Measure) score is applied to calculate the structure similarity. \textbf{(3) Text Similarity}. We apply OCR tools to recognize the text in the webpages, and then use the BLEU score \cite{papineni2002bleu} to measure the text similarity between the two webpages. \textbf{(4) Widget Match.} Widget match~\cite{GUIPilot, kevin2018design} measures the widget-level consistency between the original UI and the generated UI. We calculate the Widget Similarity (WS) and Widget Match Rate (WMS) based on the method proposed by GUIPilot~\cite{GUIPilot}.

\textbf{Interaction Part Metrics}. We also evaluate the interactive parts of webpages from the perspective of the position and function of the interaction.\textbf{(1) Position Similarity}. The position similarity between original interaction $I_o$ and generated interaction $I_g$ is defined as $P(I_o, I_g) = 1  - max(|x_o-x_g|, |y_o-y_g|)$, where $(x_o, y_o)$ and $(x_g, y_g)$ are normalized coordinates (in $[0, 1]$) of the interactive area center. 
\textbf{(2) Implement Rate (IR)} measures the ratio of interactions successfully implemented by MLLM. An interaction is considered implemented if detectable by webdriver, and unimplemented otherwise. Let $N(\cdot)$ denote the quantity, we can calculate the \textbf{IR} as $IR = \frac{N(implemented)}{N(implemented) + N(unimplemented)}$. \textbf{(3) Usability Rate (UR)}. Human annotators are asked to interact with the generated webpage and judge the usability. We can calculate as $UR = \frac{N(usable)}{N(usable) + N(unusable)}$. We also employ human annotators to conduct pair-wise comparison and failure type analysis in Section~\ref{subsec:he} and Section~\ref{subsec:failure}.






\section{Study Setup}

\subsection{Evaluation Models}


To understand the MLLMs’ performance on \task \ task and identify the gap between open-source and closed-source models, we conduct experiments on three popular commercial models: Gemini-1.5-flash \cite{google_gemini_api}, GPT-4o-20240806 \cite{openai_gpt4o} and Claude-3.5-Sonnet-20240620 \cite{anthropic2023}. \textbf{\task \ task takes multiple images as input, and many open source MLLMs do not support that (e.g., llava \cite{liu2024llavanext}, llama-3.2-vision \cite{meta_llama})}, so we select Qwen2.5-vl-instruct (3B, 7B, 72B) \cite{Qwen2.5-VL} for assessment. In configuring the MLLM models, we set the temperature to 0 and the maximum number of tokens output for Gemini-1.5-flash, GPT-4o, and Claude-3.5-Sonnect as 4096. The maximum output token for the Qwen series models is set to 2048. All other parameters were kept at their default settings as outlined in the relevant API documentation~\cite{google_gemini_api, openai_vision_guide, anthropic_vision_docs, Qwen2.5-VL}.

\subsection{Prompt Design}

\begin{figure*}[t]
    \centering
    \includegraphics[width = .99\textwidth]{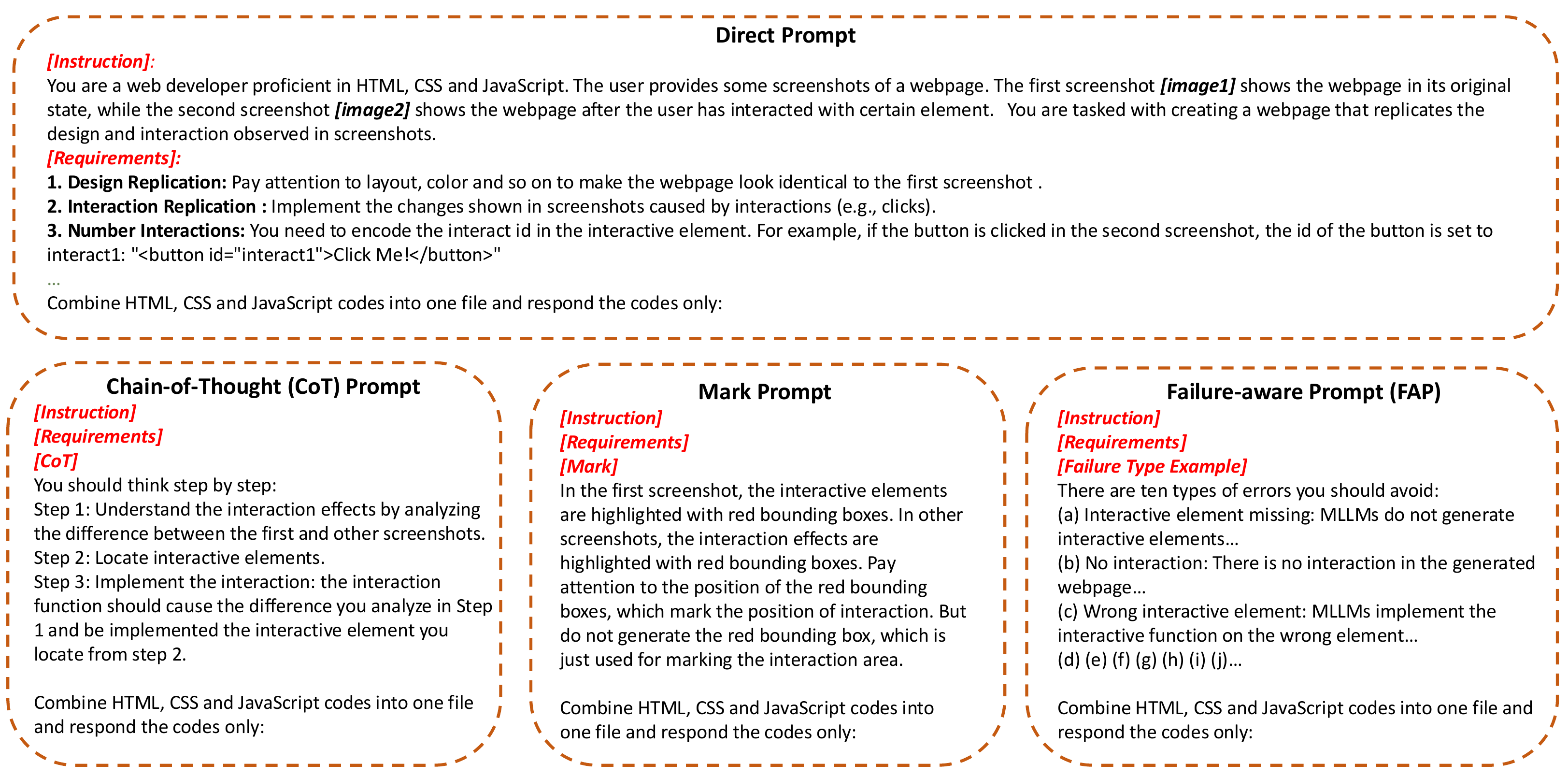}
    \caption{The four kinds of prompts for MLLMs.}
    \label{fig:direct_prompt}
\end{figure*}

The prompts are shown in Figure~\ref{fig:direct_prompt}. In the \textbf{Direct prompt}, the first screenshot represents the original webpage state, while subsequent screenshots depict states after specific interactions. Requirements are applied to guide MLLMs in replicating the webpage design and interaction.  Requirement 3 allows MLLM to number the interactions when generating code, so that in the automated testing phase, webdriver \footnote{https://selenium-python.readthedocs.io/} can locate the interactive elements through the interaction ID (e.g., interact1) and perform the interaction automatically.

To achieve Interactive element highlighting, we design CoT and Mark prompt to let MLLM focus on the interactive part. For the \textbf{CoT prompt} \cite{wei2022chain}, we use the instruction ``let’s think step by step'' and design three intermediate steps: analyze the interaction effects, locate the interactive elements, and implement the interaction. For the \textbf{Mark prompt}, we use red bounding boxes to highlight the interaction area, prompting MLLMs to focus on the interactive parts.

To enable MLLM to avoid potential errors as much as possible when generating interactions, we design \textbf{Failure-aware prompt} to put the failure types in the prompt to guide MLLM to avoid corresponding failures.

\subsection{Data Contamination Check}
Although the C4 dataset is widely used for training large models, it is unlikely that our test set was seen by MLLMs during training. The C4 dataset contains only text data, without HTML/CSS/JavaScript code, and current MLLM training datasets, like Qwen~\cite{Qwen2.5-VL}, do not include interactive prototyping data. Additionally, many existing benchmarks~\cite{si2024design2code, gui2024vision2ui} derived from the C4 dataset still present challenges for MLLMs. To confirm that MLLMs are not ``cheating'' by recognizing and reciting the original webpage code, we compute the BLEU~\cite{papineni2002bleu} score between the generated and original webpage HTML. The very low BLEU score (Table~\ref{tab:contamination}) indicates that the model generates the code based on the interactive prototype, not by directly copying the original webpage.

\begin{table}[ht]
\centering
\caption{Model Code Similarity with Original Website.}
\label{tab:contamination}
\begin{tabular}{@{}lc@{}}
\toprule
\textbf{Model}      & \textbf{BLEU Sim.} \\ \midrule
Qwen2.5-vl-72B         & 0.1274                   \\
Gemini-1.5-flash              & 0.1368                   \\
GPT-4o                 & 0.1214                   \\
Claude-3.5-Sonnet              & 0.1294                   \\ \bottomrule
\end{tabular}

\end{table}

\section{Experiments}

\subsection{Model Performance}
\label{sec:model_per}

\begin{table*}[ht]
\definecolor{darkred}{RGB}{255,0,0}
\centering
\small
\setlength{\tabcolsep}{0.5em}
\caption{Performance of different MLLMs under different prompts on \task \ task. \textbf{Bold values} indicate the optimal performance, and \underline{underlined values} indicate the second-best performance. The red value is the highest value among the averages. WS denotes the Widget Similarity, WMR denotes the widget match rate and the IR denote the implement rate.}
\label{tab:rq1_2}
\begin{tabular}{@{}c|c|ccccc|ccccc@{}}
\toprule
\multirow{2}{*}{Model} & \multicolumn{1}{c}{\multirow{2}{*}{Prompt}} & \multicolumn{5}{|c}{Full Page} & \multicolumn{5}{|c}{Interaction Part} \\ \cmidrule(l){3-12}
& \multicolumn{1}{c|}{} & CLIP & SSIM & Text & WS & WMR & CLIP & SSIM & Text & Position & IR \\ \cmidrule(r){1-12}
\multirow{4}{*}{Qwen2.5-vl-3B-instruct} & Direct & \textbf{0.3220} & \textbf{0.1932} & \textbf{0.1510} & \textbf{0.0203} & \textbf{0.3462} & \textbf{0.2100} & \textbf{0.1531} & \underline{0.0415} & \textbf{0.2090} & \textbf{0.3449} \\
& CoT & 0.2031 & 0.1085 & 0.0800 & 0.0185 & 0.3302 & 0.1219 & 0.0894 & 0.0352 & 0.1212 & 0.1979 \\
& Mark & \underline{0.2752} & \underline{0.1503} & \underline{0.1200} & \underline{0.0190} & \underline{0.3418} & \underline{0.1706} & \underline{0.1188} & \textbf{0.0514} & \underline{0.1706} & \underline{0.2647} \\
& Average & 0.2668 & 0.1507 & 0.1170 & 0.0193 & 0.3394 & 0.1675 & 0.1204 & 0.0427 & 0.1669 & 0.2692 \\ \midrule
\multirow{4}{*}{Qwen2.5-vl-7B-instruct} & Direct & \underline{0.4169} & \underline{0.2886} & \underline{0.2519} & 0.0222 & \underline{0.3887} & \underline{0.3230} & \underline{0.2177} & \underline{0.0952} & \underline{0.2529} & \underline{0.4786} \\
& CoT & 0.3895 & 0.2529 & 0.2207 & \textbf{0.0286} & \textbf{0.3909} & 0.2806 & 0.1981 & 0.0744 & 0.2259 & 0.4305 \\
& Mark & \textbf{0.4586} & \textbf{0.3282} & \textbf{0.2703} & \underline{0.0234} & 0.3666 & \textbf{0.3541} & \textbf{0.2468} & \textbf{0.1348} & \textbf{0.2798} & \textbf{0.5267} \\
& Average & 0.4217 & 0.2899 & 0.2477 & 0.0247 & 0.3821 & 0.3192 & 0.2209 & 0.1015 & 0.2529 & 0.4786 \\ \midrule
\multirow{4}{*}{Qwen2.5-vl-72B-instruct} & Direct & \underline{0.6430} & 0.4234 & 0.4197 & \underline{0.0371} & 0.4285 & 0.4624 & 0.3207 & \underline{0.2450} & 0.3950 & 0.6524 \\
& CoT & 0.6335 & \textbf{0.4785} & \underline{0.4585} & 0.0369 & \underline{0.4395} & \textbf{0.5090} & \textbf{0.3692} & 0.2376 & \underline{0.4385} & \textbf{0.7380} \\
& Mark & \textbf{0.6954} & \underline{0.4569} & \textbf{0.4586} & \textbf{0.0401} & \textbf{0.4430} & \underline{0.4992} & \underline{0.3621} & \textbf{0.2995} & \textbf{0.4541} & \underline{0.7112} \\
& Average & 0.6573 & 0.4529 & 0.4456 & 0.0380 & 0.4370 & 0.4902 & 0.3507 & 0.2607 & 0.4292 & 0.7005 \\ \midrule
\multirow{4}{*}{Gemini-1.5-flash} & Direct & 0.5967 & 0.4526 & 0.4749 & \underline{0.0330} & \textbf{0.4964} & 0.4737 & 0.3616 & 0.2809 & 0.4320 & 0.6738 \\
& CoT & \underline{0.6166} & \underline{0.4810} & \underline{0.4775} & \textbf{0.0332} & 0.4783 & \underline{0.5093} & \underline{0.3854} & \underline{0.3217} & \underline{0.4511} & \underline{0.7112} \\
& Mark & \textbf{0.6321} & \textbf{0.4946} & \textbf{0.4878} & 0.0322 & \underline{0.4826} & \textbf{0.5194} & \textbf{0.3898} & \textbf{0.3454} & \textbf{0.4612} & \textbf{0.7326} \\
& Average & 0.6151 & 0.4761 & 0.4801 & 0.0328 & 0.4858 & 0.5008 & 0.3789 & 0.3160 & 0.4481 & 0.7059 \\ \cmidrule(r){1-12}
\multirow{4}{*}{GPT-4o} & Direct & \underline{0.7114} & \underline{0.5277} & \textbf{0.5147} & \textbf{0.0440} & \underline{0.4645} & \underline{0.5605} & \underline{0.4149} & 0.3590 & \underline{0.4888} & \underline{0.7754} \\
& CoT & 0.6905 & 0.4962 & 0.4761 & \underline{0.0435} & \textbf{0.4674} & 0.5234 & 0.4013 & \underline{0.3663} & 0.4668 & 0.7273 \\
& Mark & \textbf{0.7160} & \textbf{0.5539} & \underline{0.5112} & 0.0414 & 0.4418 & \textbf{0.5955} & \textbf{0.4488} & \textbf{0.4474} & \textbf{0.5225} & \textbf{0.8128} \\
& Average & 0.7059 & \textcolor{darkred}{0.5259} & 0.5007 & 0.0430 & 0.4579 & 0.5598 & \textcolor{darkred}{0.4217} & 0.3909 & 0.4927 & 0.7718 \\ \cmidrule(r){1-12}
\multirow{4}{*}{Claude-3.5-Sonnet} & Direct & \underline{0.7172} & \textbf{0.5318} & \textbf{0.6003} & \textbf{0.0533} & \textbf{0.5284} & \underline{0.5674} & \underline{0.4209} & \underline{0.3833} & \underline{0.5123} & \underline{0.7914} \\
& CoT & 0.6961 & 0.5110 & 0.5603 & 0.0451 & 0.5099 & 0.5606 & 0.4005 & 0.3662 & 0.5085 & 0.7727 \\
& Mark & \textbf{0.7258} & \underline{0.5299} & \underline{0.5899} & \underline{0.0486} & \underline{0.5111} & \textbf{0.5944} & \textbf{0.4282} & \textbf{0.4319} & \textbf{0.5149} & \textbf{0.7968} \\
& Average & \textcolor{darkred}{0.7130} & 0.5242 & \textcolor{darkred}{0.5835} & \textcolor{darkred}{0.0490} & \textcolor{darkred}{0.5165} & \textcolor{darkred}{0.5742} & 0.4165 & \textcolor{darkred}{0.3938} & \textcolor{darkred}{0.5119} & \textcolor{darkred}{0.7870} \\ \bottomrule
\end{tabular}
\end{table*}

\subsubsection{Automatic Evaluation}

We evaluate the performance of MLLMs on the Interaction2Code task using the metrics outlined in Section~\ref{subsec:evaluation}. The results are presented in Table~\ref{tab:rq1_2}. Our observations are as follows: (1) GPT-4o and Claude-3.5-Sonnet outperform other models on average. (2) Among open-source models, Qwen2.5-vl-72B achieves the best performance, comparable to the commercial model Gemini-1.5-flash. Performance improves as model size increases. (3) \textbf{MLLMs perform worse on the interactive part compared to the full page (Limitation 1).} This limitation arises from the MLLMs' insufficient focus on the interactive component, motivating our proposed solution to prioritize the interaction part.

\begin{tcolorbox}[colback=gray!20, colframe=gray!20, width=\columnwidth]
\textbf{Improvement 1: Interactive element highlighting.} To improve the performance of generated interaction, we further propose \textit{Chain-of-Thought (CoT) and Mark prompts} to force models to focus on the interaction.
\end{tcolorbox}

For the CoT prompt \cite{wei2022chain}, we design three thinking steps: analyze the interaction effects, locate the interactive elements, and implement the interaction. For the Mark prompt, we highlight the interaction area with red bounding boxes to direct MLLMs' focus on the interaction.

\textbf{Both CoT and Mark prompts improve model performance compared to direct prompting, with the Mark prompt showing superior results.} For Gemini-1.5-flash, the metrics (CLIP, SSIM, text, position, IR) for the interaction part improve from the direct prompt scores (0.4737, 0.3616, 0.2809, 0.4302, 0.6738) to (0.5093, 0.3854, 0.3217, 0.4511, 0.7112) with CoT, and further to (0.5194, 0.3898, 0.3454, 0.4612, 0.7326) with the Mark prompt.


\textbf{The widget similarity scores demonstrate consistently poor performance.} This finding underscores fundamental limitations in the model's capacity to accurately interpret widget characteristics and hierarchical layout structures. Furthermore, analysis of GUIPilot's~\cite{GUIPilot} output reveals significant omissions of web page elements. This stems from the fact that existing design consistency methodologies~\cite{GUIPilot, kevin2018design} are primarily developed for mobile UI contexts, where interface images are relatively compact and uniform. In contrast, web UI images typically exhibit greater complexity and scale, thereby presenting substantial challenges to widget detection algorithms, thus designing the widget-level consistency detection algorithm suitable for webpage is necessary.

\subsubsection{Human Evaluation}
\label{subsec:he}


\begin{figure}[t]
    \subfigure[Usability evaluation.]{
    \label{fig:compare1}
    \centering
    \includegraphics[width = .22\textwidth]{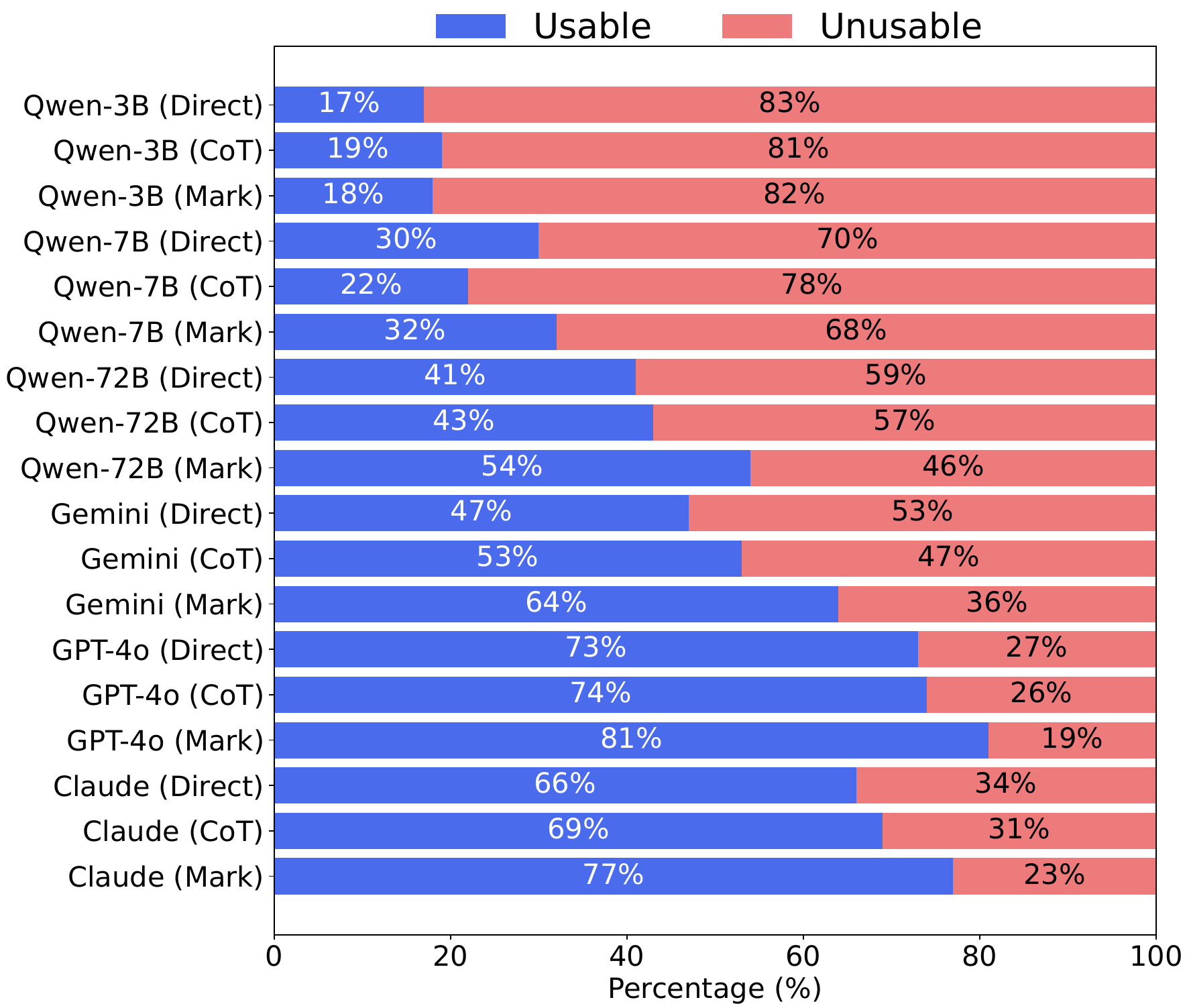}
    }
    \subfigure[Pairwise comparision.]{
    \label{fig:compare}
    \centering
    \includegraphics[width = .22\textwidth]{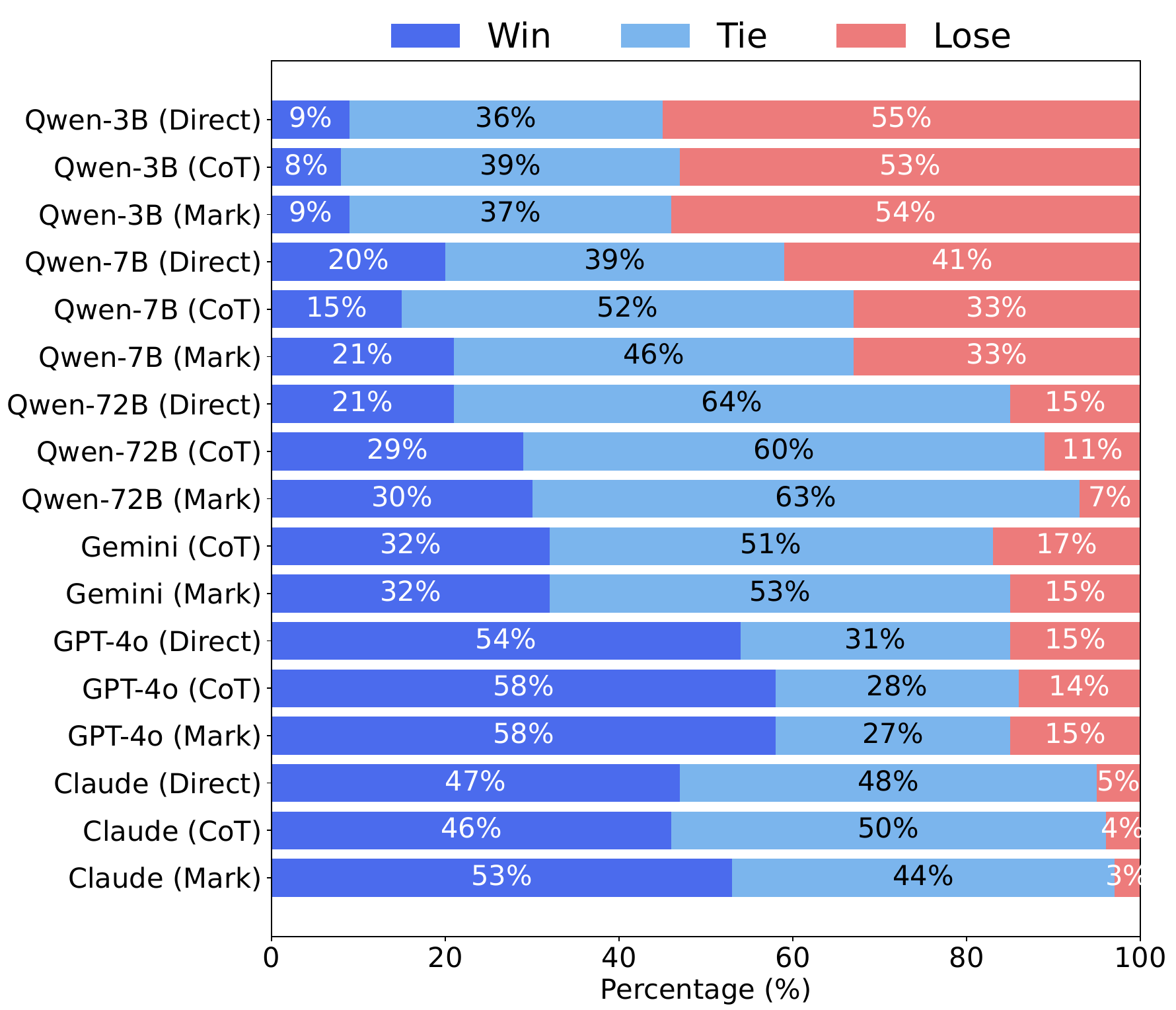}
    }
    \caption{Human evaluation, a higher usable rate indicates better functionality and a higher win rate indicates better quality.}
\end{figure}

 \paragraph{Functionality Evaluation} In addition to automatic metrics, we conduct a functionality evaluation with four PhD students, each with three years of front-end development experience, to assess the usability of the generated interactions. An interaction is considered usable if its implementation precisely matches the behavior specified in the interactive prototype. The usability rate results are shown in Figure~\ref{fig:compare1}.

\paragraph{Pairwise Model Comparison} We ask five human annotators to rank a pair of generated interactions (one from the baseline, the other from the tested methods) to decide which one implements the reference interaction function better. We use Gemini-1.5-flash with direct prompt as the baseline and collect the other 17 methods’ Win/Tie/Lose rates against this baseline. Each pair will count as Win (Lose) only when Win (Lose) receives the majority vote ($\geq 3$). All other cases are considered a Tie. The results are shown in Figure~\ref{fig:compare}, where a higher win rate and lower loss rate suggest better quality as judged by human annotators.

\textbf{Results.} (1) Our human evaluation reveals that GPT-4o and Claude-3.5-Sonnet consistently outperform other baseline models. (2) Both CoT and Mark prompting strategies can enhance model performance beyond direct prompting, showing higher win rates and usability rates across most models (except Qwen-vl-7B-instruct's CoT prompt). (3) Mark prompting yields the most significant improvements in usability, with Claude-3.5-Sonnet showing 11\% and 8\% increases compared to Direct and CoT prompts, respectively (Figure~\ref{fig:compare1}). (4) These human evaluation results are consistent with Section~\ref{sec:model_per}, confirming the validity of our automatic evaluation metrics. Detailed instructions for human evaluations are available in our artifact.

\subsection{Failure Type Analysis}
\label{subsec:failure}

\begin{table*}[t]
\centering
\small
\setlength{\tabcolsep}{0.4em}
\caption{Failure types and their influences, where \redcross \ represents full impact and \partialcheck \ represents partial impact.}
\label{tab:failure}
\begin{tabular}{@{}c|c|ccc|cc@{}}
\toprule
\begin{tabular}[c]{@{}c@{}} Failure \\ Object\end{tabular}                                                               & Failure Type                       & Content & Function & \begin{tabular}[c]{@{}c@{}} User \\ Experience\end{tabular} & Usability Rate \\ \midrule
\multirow{4}{*}{\begin{tabular}[c]{@{}c@{}}Interactive\\ element\end{tabular}} & (a) Interactive element missing        &  \redcross       & \redcross           & \redcross & 0\%    \\
                                                                               & (b) No interaction                     &   \partialcheck    &    \redcross       & \redcross     & 6.93\%           \\
                                                                               & (c) Wrong interactive element          & \partialcheck       & \partialcheck          & \redcross   & 92.31\%            \\ 
                                                                                & (d) Wrong type of interactive element          & \partialcheck       & \partialcheck          & \redcross   & 96.82\%            \\ 
                                                                               & (e) Wrong position of interactive element & \partialcheck & \partialcheck & \redcross &  98.41\% \\
                                                                               \cmidrule(l){1-7} 
\multirow{4}{*}{\begin{tabular}[c]{@{}c@{}}Interaction\\ effects\end{tabular}} & (f) Wrong position after interaction   & \partialcheck       & \partialcheck          & \redcross    & 96.17\%           \\
                                                                               & (g) Wrong type of interaction effects & \partialcheck       & \partialcheck          & \redcross & 57.14\%             \\
                                                                               & (h) Effect on wrong element                     & \partialcheck       & \partialcheck          & \redcross & 44.44\% & \\
                                                                               & (i) Partial Implementation             & \partialcheck       & \partialcheck          & \redcross   &89.20\%       \\
                                                                               & (j) Wrong function                     & \redcross       & \redcross          & \redcross      & 0\%               \\ \bottomrule
\end{tabular}

\end{table*}

\begin{figure*}[ht]
    \centering
    \includegraphics[width = .95\textwidth]{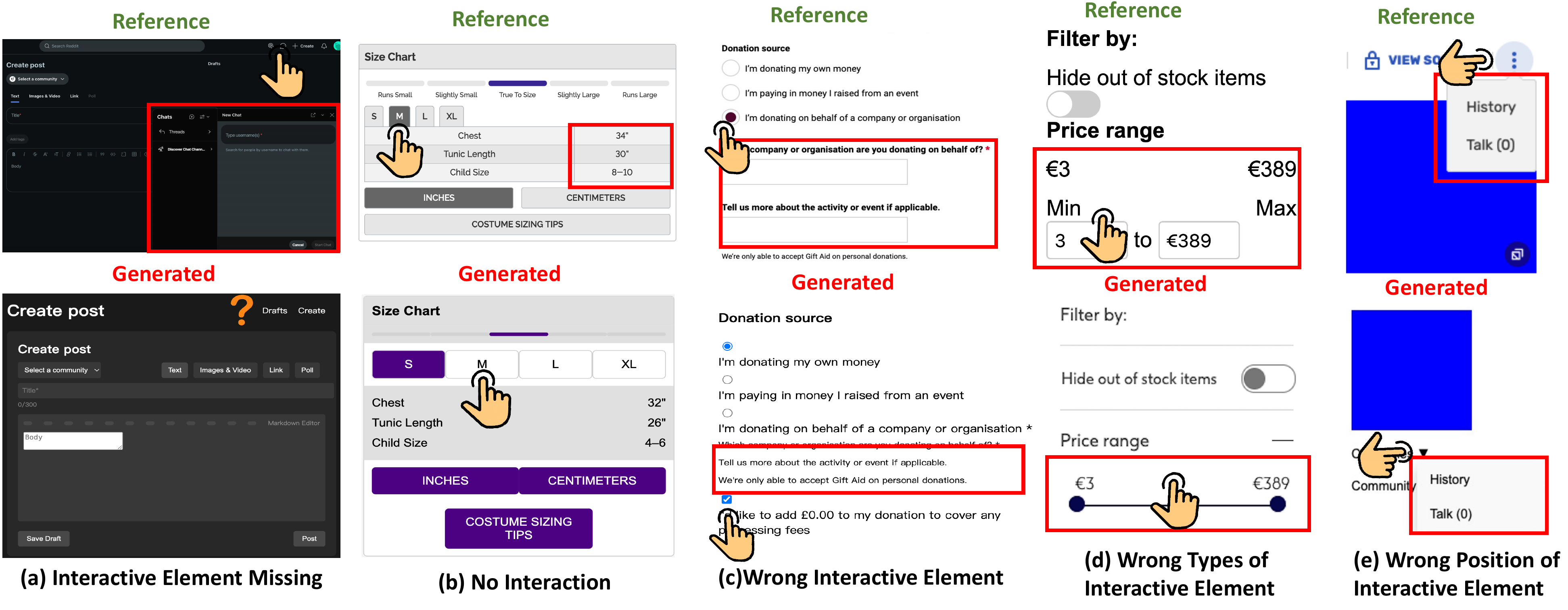}
    \caption{Failure on interactive elements.}
    \label{fig:error1}
\end{figure*}

\begin{figure*}[ht]
    \centering
    \includegraphics[width = .95\textwidth]{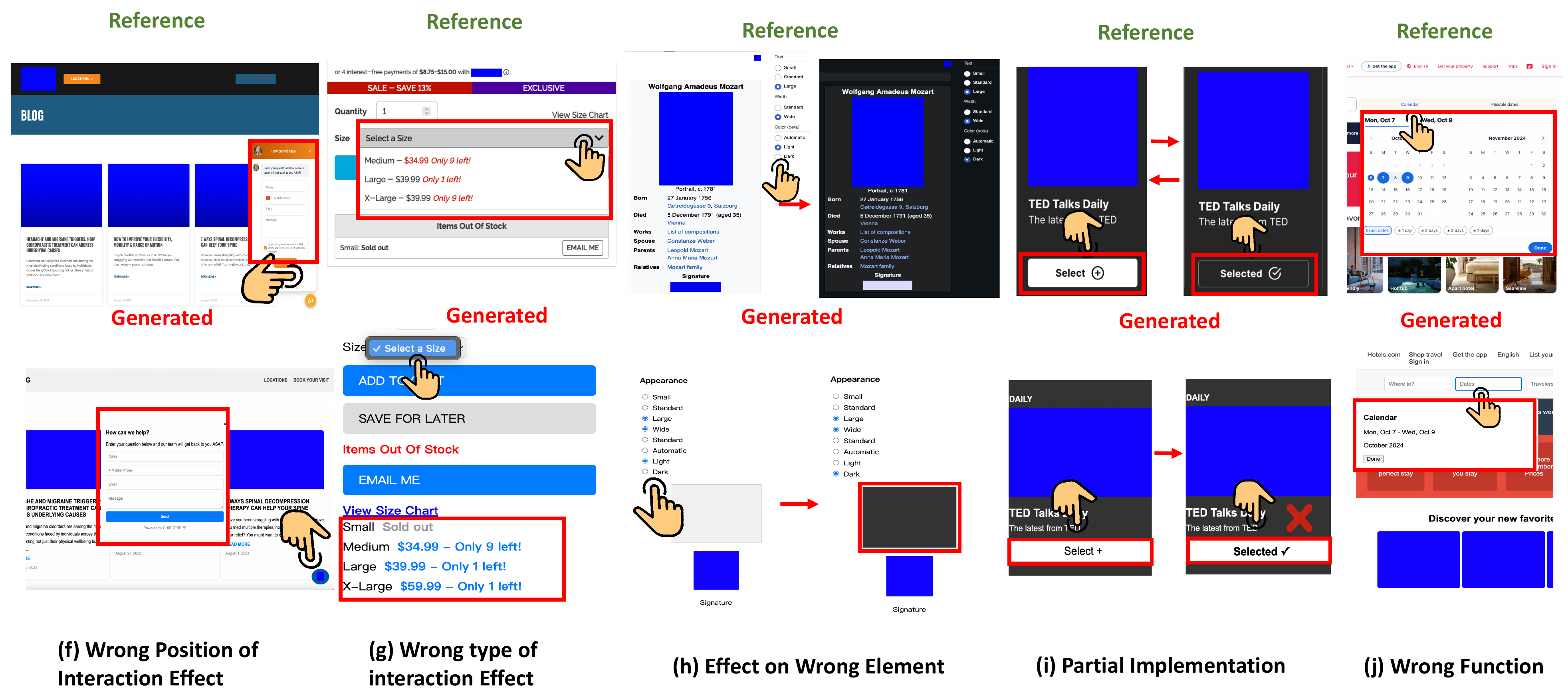}
    \caption{Failure of interaction effects.}
    \label{fig:error2}
\end{figure*}

To analyze the differences between the generated and original interactions, categorize the failure types, and assess their impact on content, functionality, and user experience, we employ four PhD students with three years of front-end development experience. We first randomly select 25\% interactions for analysis and then discuss, revise, and refine the failure type until everyone reaches a consensus. During annotation, if the annotators encounter a new failure type, they will communicate and update the failure types in time to guide subsequent annotations (the detailed instructions are available in our artifact). The results are in Table~\ref{tab:failure}, which shows that \textbf{MLLMs are prone to make 10 types of failure (Limitation 2)}. Figure~\ref{fig:error1} and Figure~\ref{fig:error2} demonstrate examples of these failures, where the first row shows the reference interaction, and the second row shows the generated interaction by MLLMs. The failures are illustrated as follows:

\subsubsection{Failure on Interactive Elements}

\begin{enumerate}[label=(\alph*)]
    \item Missing interactive element: MLLMs fail to generate interactive elements. For example, in Figure~\ref{fig:error1}(a), the reference webpage has a chat button that opens a chat window when clicked, but the generated webpage lacks this button, preventing any interaction.
    \item No interaction: The generated webpage lacks interactive behavior. For instance, in Figure~\ref{fig:error1}(b), clicking the "M" button on the reference webpage changes the displayed size, but clicking it in the generated page produces no change. 
    \item Wrong interactive element: MLLMs implement the interactive function on the wrong element. For instance, in Figure~\ref{fig:error1}(c), the original webpage displays input boxes after clicking "I'm donating on behalf of a company," but the generated webpage shows them only after clicking a different element.
    \item Wrong type of interactive element: MLLMs generate the wrong type of interactive element. For example, in Figure~\ref{fig:error1}(d), the price adjustment element in the original page is an input field, while in the generated page, it is a progress bar.
    \item Wrong position of interactive element: MLLMs place interactive elements in the wrong position. In Figure~\ref{fig:error1}(e), the button on the reference webpage is in the upper-right corner, but in the generated page, it is positioned below the image.
\end{enumerate}

\subsubsection{Failure on Interactive Effects}

\begin{enumerate}[label=(\alph*), start=6]
    \item Wrong position after interaction: The interaction effects appear in the wrong position. For example, in Figure~\ref{fig:error2}(f), the reference webpage shows a pop-up window in the lower-left corner after clicking the dialogue button, but the generated webpage displays it in the center.
    \item Wrong type of interaction effects: In Figure~\ref{fig:error2}(g), the reference webpage shows an option-type element after clicking select, while the generated page shows a text-type element.
    \item Effect on wrong element: MLLMs apply the interaction effect to the wrong element. As seen in Figure~\ref{fig:error2}(h), in the reference webpage, clicking the "dark" button changes the background color to black, but in the generated page, only a block turns black, leaving the background unchanged.
    \item Partial Implementation: MLLMs only implement part of the interactive functionality. For instance, in Figure~\ref{fig:error2}(i), the reference webpage allows a button to be selected and unselected, but the generated webpage only allows the button to be selected, not unselected.
    \item Wrong function: MLLMs implement the wrong function. As shown in Figure~\ref{fig:error2}(j), in the original webpage, clicking the button opens a date selection box, but in the generated webpage, it opens a date display box instead.
\end{enumerate}



\textbf{Failure reason analysis.} Specifically, we cluster 10 failure types into three main symptoms that all related to the above limitation:(1) Spatial failures (Failures a, c, e, f) – the model cannot accurately locate the element linked to the interaction. (2) Type failures (Failures d, g) – it misclassifies the widget type (e.g., slider vs. input), resulting in incorrect HTML elements in the output code.
(3) Behavioral failures (Failures b, h, i, j) – it fails to infer the correct state-change logic (e.g., toggle vs. open-modal), leading to missing or incorrect code implementation.

\begin{figure*}[t]
    \subfigure[Qwen2.5-vl-3B.]{
    \label{fig:qwen3}
    \centering
    \includegraphics[width = .3\textwidth]{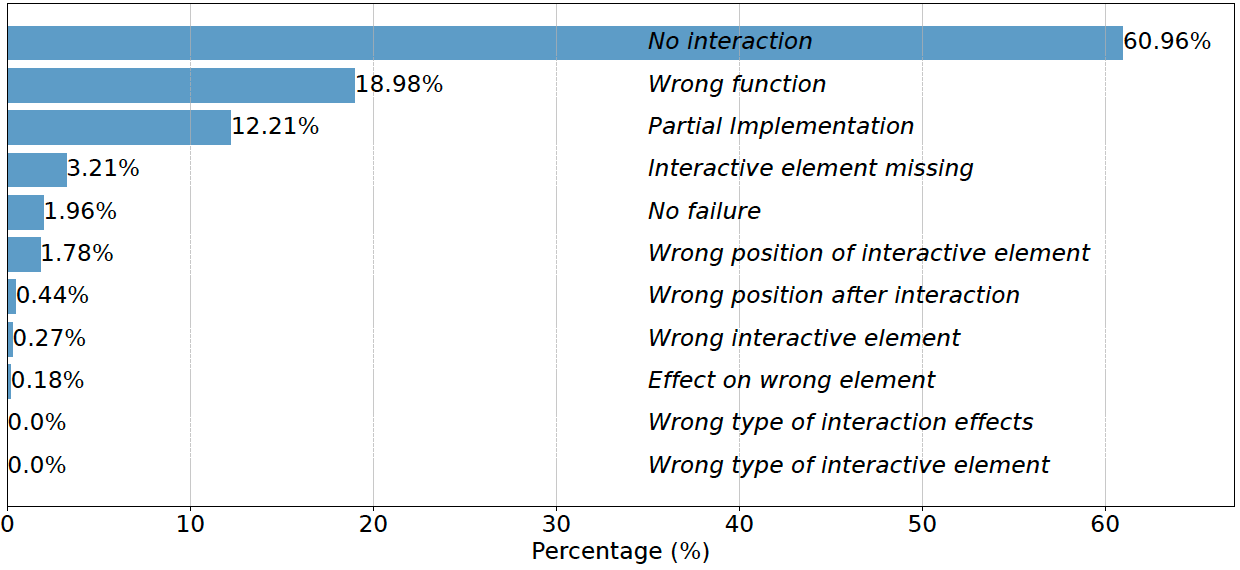}
    }
    \subfigure[Qwen2.5-vl-7B.]{
    \label{fig:qwen7}
    \centering
    \includegraphics[width = .3\textwidth]{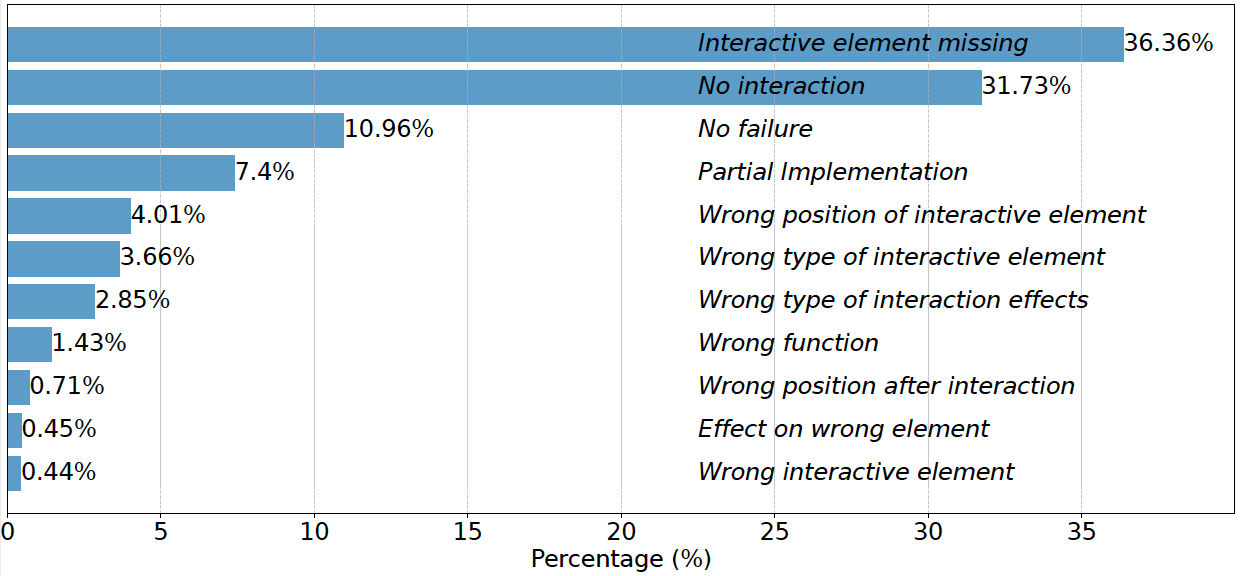}
    }
    \subfigure[Qwen2.5-vl-72B.]{
    \label{fig:qwen72}
    \centering
    \includegraphics[width = .3\textwidth]{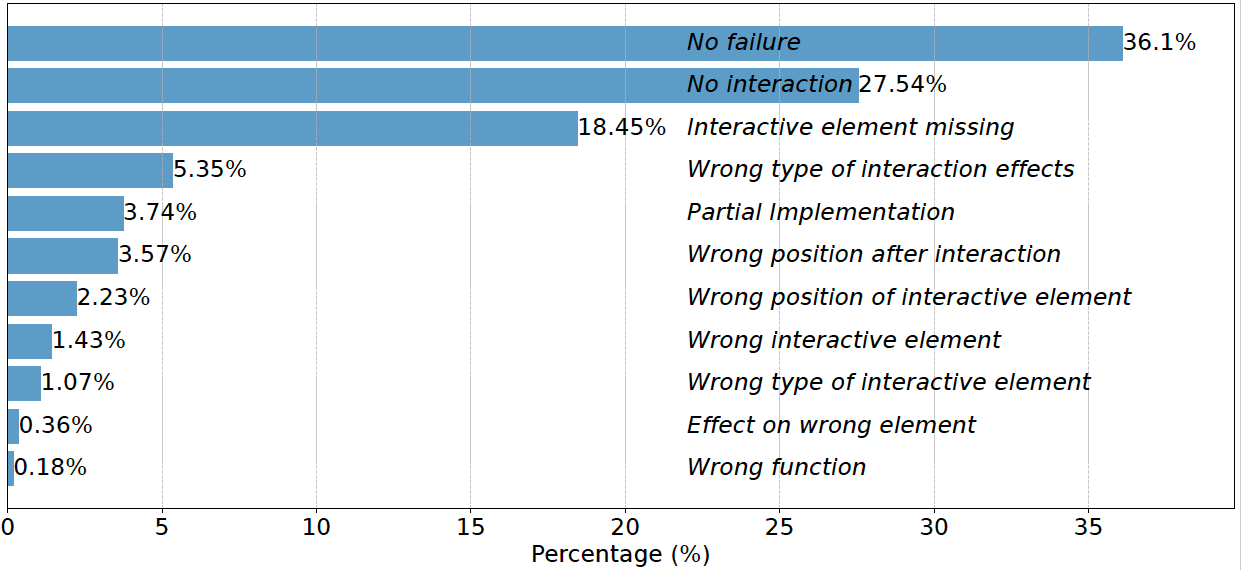}
    }
    \subfigure[Gemini-1.5-flash.]{
    \label{fig:gemini0}
    \centering
    \includegraphics[width = .3\textwidth]{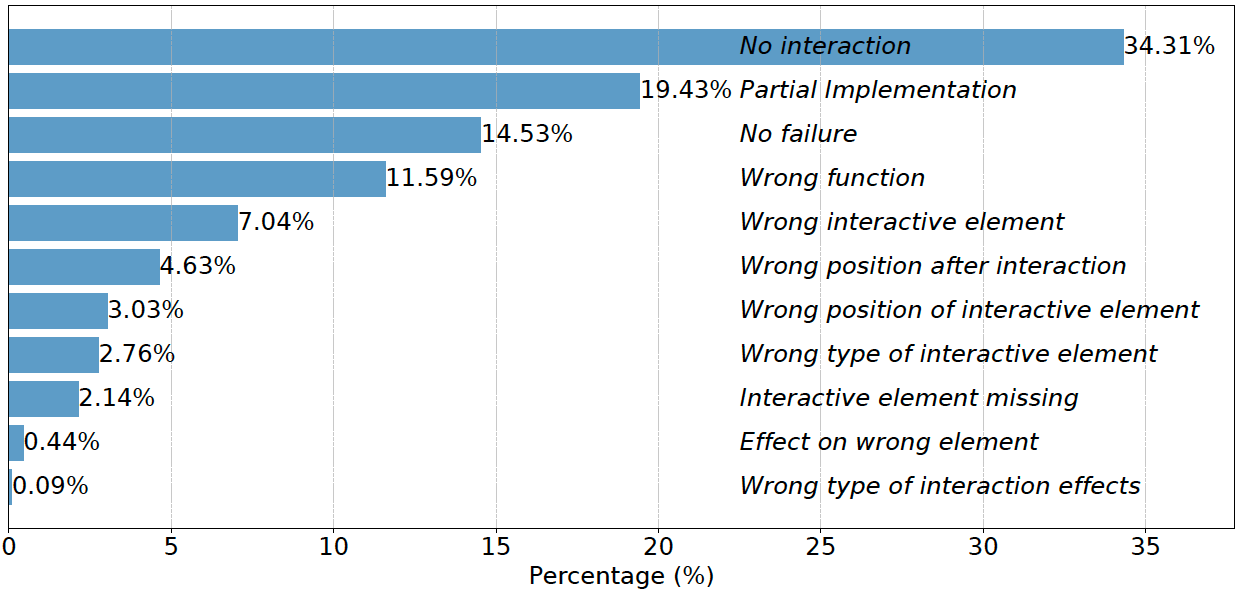}
    }\hspace{0.3cm}
    \subfigure[GPT-4o.]{
    \label{fig:gpt0}
    \centering
    \includegraphics[width = .3\textwidth]{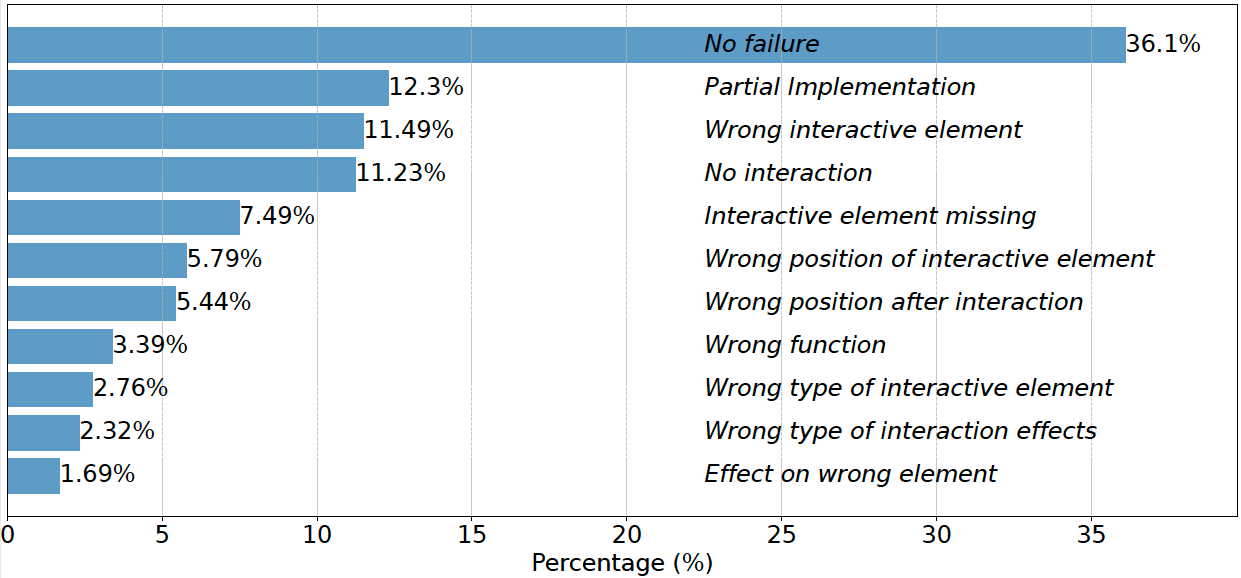}
    }\hspace{0.4cm}
    \subfigure[Claude-3.5-Sonnet.]{
    \label{fig:claude0}
    \centering
    \includegraphics[width = .3\textwidth]{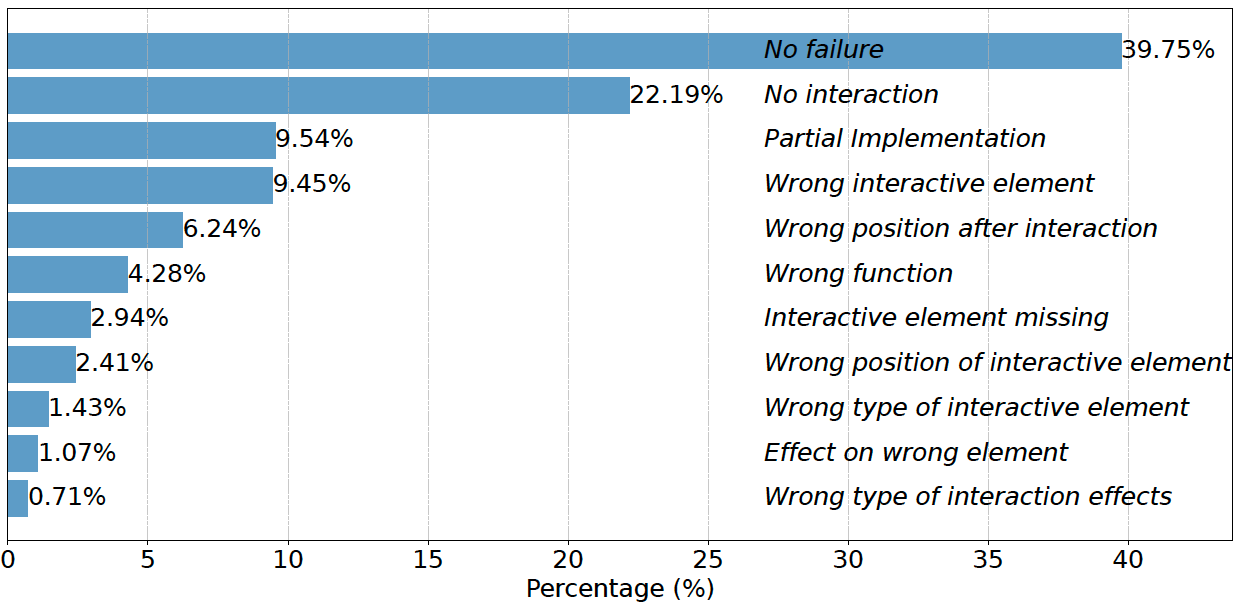}
    }
    \caption{Failure distribution of MLLMs.}
    \label{fig:failure_dis}
\end{figure*}

Based on the failure distribution in Figure~\ref{fig:failure_dis}, we find that, \textbf{the main failure modes are ``No interaction'', ``Partial implementation'', ``Interactive element missing'', and ``Wrong function''.}

\textbf{Additionally, the most critical failures are ``Interactive element missing'', ``Wrong function'', ``No interaction'' and ``Effect on wrong element''.} The severity of these failures is determined by the usability rate (UR), where a higher UR indicates lower severity, and a lower UR indicates higher severity. As illustrated in Table~\ref{tab:failure}, failures (a), (b), and (j) exhibit a UR lower than 10\%, rendering the generated interactions completely ineffective.


\begin{tcolorbox}[colback=gray!20, colframe=gray!20, width=\columnwidth]
\textbf{Improvement 2: Failure-aware Prompt (FAP).} Based on failure types, we propose FAP to stimulate the self-criticism ability of MLLM, thereby avoiding problems that may occur in the \task \ task.
\end{tcolorbox}
FAP incorporates the failure example into the prompt and tell MLLMs to avoid these types of failures (as shown in Figure~\ref{fig:direct_prompt} ). We use $\frac{2}{3}$ of the dataset to annotate failure types and $\frac{1}{3}$ of the dataset to test. Table~\ref{tab:critic} shows the results of the FAP methods, we can find that \textbf{Failure-aware Prompt can improve the performance of the \task \ task on all models.} 


\begin{table*}[ht]
\centering
\definecolor{darkgreen}{RGB}{0,150,0}
\definecolor{darkred}{RGB}{200,0,0}
\setlength{\tabcolsep}{0.4em}
\caption{Comparison between direct prompt and FAP.}
\label{tab:critic}
{\fontsize{10pt}{10pt}\selectfont
\begin{tabular}{@{}c|c|lll|lllll@{}}
\toprule
\multirow{2}{*}{Model} & \multirow{2}{*}{Method} & \multicolumn{3}{c|}{Full Page} & \multicolumn{5}{c}{Interaction Part} \\
\cmidrule(l){3-10}
& & CLIP & SSIM & Text & CLIP & SSIM & Text & Position & IR \\ \midrule
\multirow{3}{*}{Gemini-1.5-flash} & Direct & 0.6276 & 0.4984 & 0.5231 & 0.5403 & 0.4494 & 0.3602 & 0.5802 & 0.7636 \\
& FAP & 0.6580 & 0.5337 & 0.5311 & 0.5886 & 0.4584 & 0.4394 & 0.6032 & 0.8182 \\
& $\Delta$ & \textcolor{darkgreen}{$\uparrow$ 0.0304} & \textcolor{darkgreen}{$\uparrow$ 0.0353} & \textcolor{darkgreen}{$\uparrow$ 0.0080} & \textcolor{darkgreen}{$\uparrow$ 0.0483} & \textcolor{darkgreen}{$\uparrow$ 0.0090} & \textcolor{darkgreen}{$\uparrow$ 0.0792} & \textcolor{darkgreen}{$\uparrow$ 0.0230} & \textcolor{darkgreen}{$\uparrow$ 0.0546} \\ \midrule
\multirow{3}{*}{GPT-4o} & Direct & 0.6660 & 0.5480 & 0.4995 & 0.5700 & 0.4891 & 0.3652 & 0.5803 & 0.7636 \\
& FAP & 0.7047 & 0.5976 & 0.6045 & 0.6072 & 0.5405 & 0.4580 & 0.6452 & 0.8364 \\
& $\Delta$ & \textcolor{darkgreen}{$\uparrow$ 0.0387} & \textcolor{darkgreen}{$\uparrow$ 0.0496} & \textcolor{darkgreen}{$\uparrow$ 0.1050} & \textcolor{darkgreen}{$\uparrow$ 0.0372} & \textcolor{darkgreen}{$\uparrow$ 0.0514} & \textcolor{darkgreen}{$\uparrow$ 0.0928} & \textcolor{darkgreen}{$\uparrow$ 0.0649} & \textcolor{darkgreen}{$\uparrow$ 0.0728} \\ \midrule
\multirow{3}{*}{Claude-3.5-Sonnet} & Direct & 0.5747 & 0.3950 & 0.4611 & 0.4582 & 0.3771 & 0.3086 & 0.4927 & 0.6364 \\
& FAP & 0.6080 & 0.4500 & 0.4810 & 0.4921 & 0.4035 & 0.3822 & 0.5154 & 0.6545 \\
& $\Delta$ & \textcolor{darkgreen}{$\uparrow$ 0.0333} & \textcolor{darkgreen}{$\uparrow$ 0.0550} & \textcolor{darkgreen}{$\uparrow$ 0.0199} & \textcolor{darkgreen}{$\uparrow$ 0.0339} & \textcolor{darkgreen}{$\uparrow$ 0.0264} & \textcolor{darkgreen}{$\uparrow$ 0.0736} & \textcolor{darkgreen}{$\uparrow$ 0.0227} & \textcolor{darkgreen}{$\uparrow$ 0.0181} \\ \midrule
\multirow{3}{*}{Qwen2.5-vl-3B-instruct} & Direct & 0.4284 & 0.2466 & 0.1674 & 0.2777 & 0.2180 & 0.0285 & 0.3020 & 0.4727 \\
& FAP & 0.3647 & 0.2076 & 0.1146 & 0.2375 & 0.1867 & 0.0328 & 0.2213 & 0.3818 \\
& $\Delta$ & \textcolor{darkred}{$\downarrow$ 0.0637} & \textcolor{darkred}{$\downarrow$ 0.0390} & \textcolor{darkred}{$\downarrow$ 0.0528} & \textcolor{darkred}{$\downarrow$ 0.0402} & \textcolor{darkred}{$\downarrow$ 0.0313} & \textcolor{darkgreen}{$\uparrow$ 0.0043} & \textcolor{darkred}{$\downarrow$ 0.0807} & \textcolor{darkred}{$\downarrow$ 0.0909} \\ \midrule
\multirow{3}{*}{Qwen2.5-vl-7B-instruct} & Direct & 0.3596 & 0.1981 & 0.1758 & 0.2802 & 0.1894 & 0.0854 & 0.2580 & 0.4000 \\
& FAP & 0.3828 & 0.1642 & 0.1948 & 0.2603 & 0.1747 & 0.0746 & 0.2419 & 0.4182 \\
& $\Delta$ & \textcolor{darkgreen}{$\uparrow$ 0.0232} & \textcolor{darkred}{$\downarrow$ 0.0339} & \textcolor{darkgreen}{$\uparrow$ 0.0190} & \textcolor{darkred}{$\downarrow$ 0.0199} & \textcolor{darkred}{$\downarrow$ 0.0147} & \textcolor{darkred}{$\downarrow$ 0.0108} & \textcolor{darkred}{$\downarrow$ 0.0161} & \textcolor{darkgreen}{$\uparrow$ 0.0182} \\ \midrule
\multirow{3}{*}{Qwen2.5-vl-72B-instruct} & Direct & 0.6169 & 0.3967 & 0.4060 & 0.4741 & 0.3612 & 0.3275 & 0.5022 & 0.6545 \\
& FAP & 0.6194 & 0.4208 & 0.4426 & 0.5144 & 0.3750 & 0.3286 & 0.5376 & 0.7636 \\
& $\Delta$ & \textcolor{darkgreen}{$\uparrow$ 0.0025} & \textcolor{darkgreen}{$\uparrow$ 0.0241} & \textcolor{darkgreen}{$\uparrow$ 0.0366} & \textcolor{darkgreen}{$\uparrow$ 0.0403} & \textcolor{darkgreen}{$\uparrow$ 0.0138} & \textcolor{darkgreen}{$\uparrow$ 0.0011} & \textcolor{darkgreen}{$\uparrow$ 0.0354} & \textcolor{darkgreen}{$\uparrow$ 0.1091} \\
\bottomrule
\end{tabular}}
\end{table*}

\subsection{The Impact of Interaction Visual Saliency}


The visual perception limitations of MLLMs affect their performance on visual understanding tasks, especially when facing small low-resolution objects  \cite{zhang2024exploring}. We examine the impact of interaction area ratio (i.e., visual saliency) on generation outcomes. Let $I$ denote interaction, $S_I$ denote the screenshot of the webpage after interaction $I$, we define the visual saliency 
$VS(I) = \frac{area(I)}{area(S_I)}$, where $area()$ calculates the size (in pixels) of a component. A higher VS score indicates a larger area influenced by the interaction and, consequently, a higher visual saliency.



\begin{figure*}[ht]
    \centering
    \subfigure[Visual saliency distribution.]{
        \label{fig:vs_dis}
        \includegraphics[width=0.3\textwidth]{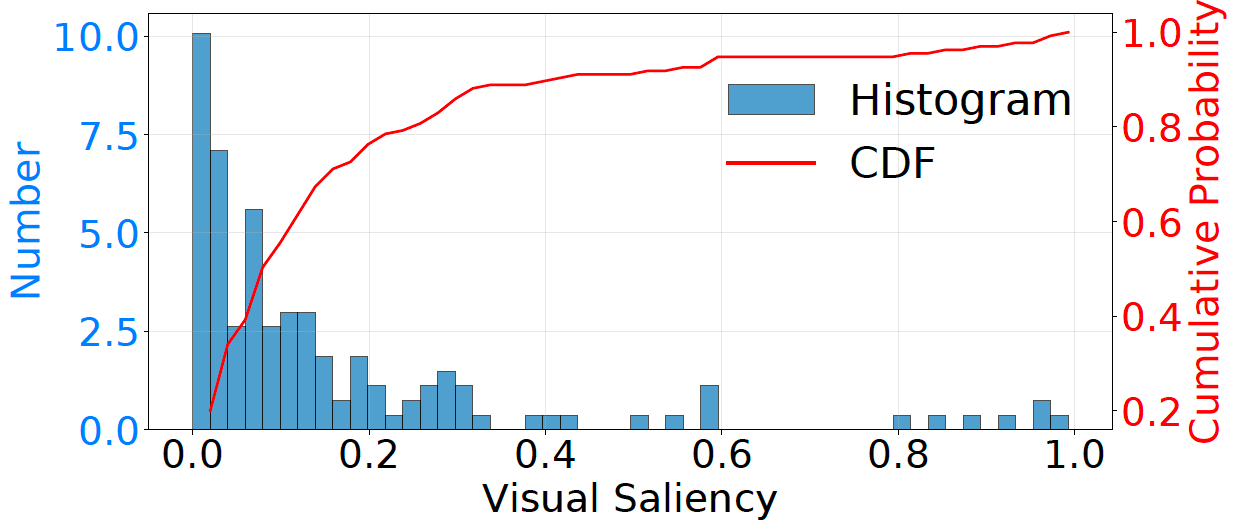}
    }
    \subfigure[CLIP and Text.]{
        \label{fig:box1}
        \includegraphics[width=0.32\textwidth]{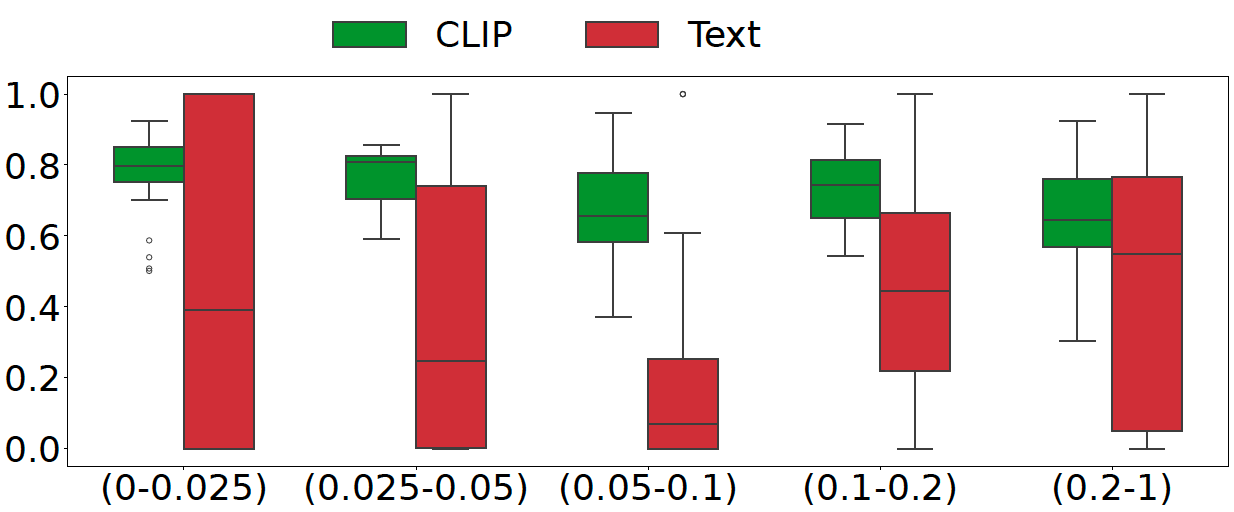}
    }
    \subfigure[SSIM and Position.]{
        \label{fig:box2}
        \includegraphics[width=0.32\textwidth]{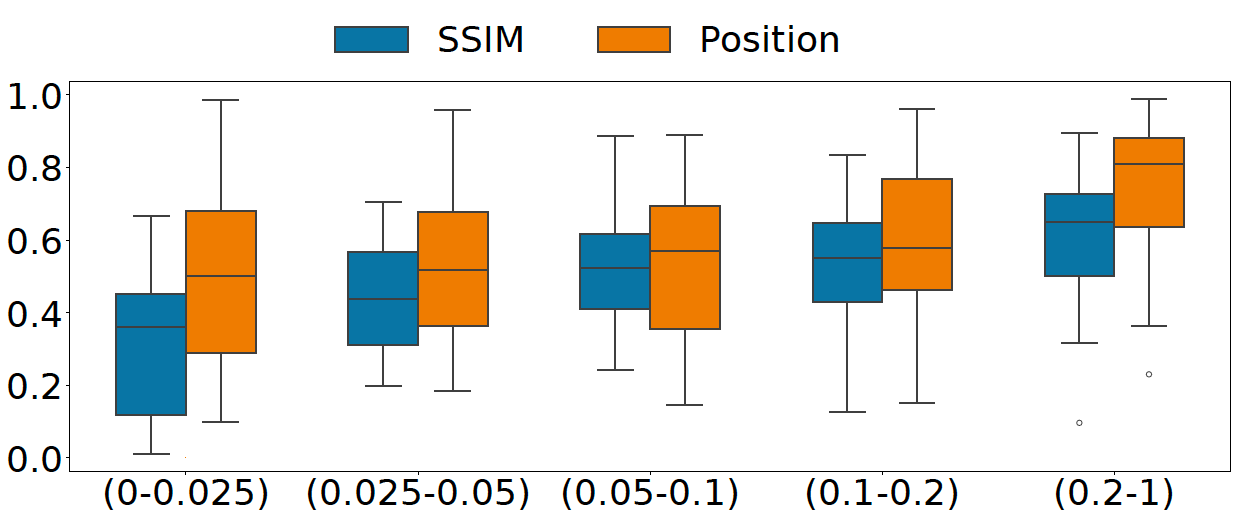}
    }
        \caption{Visual saliency and interaction part metrics distribution of different groups of Gemini-1.5-flash.}
    \label{fig:all}
\end{figure*}




We first calculate the visual saliency for all interactions and plot the distribution, as shown in Figure~\ref{fig:vs_dis}. We then divide the samples into five groups based on the distribution results, keeping the number of samples in each group roughly balanced. The VS ranges for the five groups are as follows: [0, 0.025), [0.025, 0.05), [0.05, 0.1], [0.1, 0.2), [0.2, 1). Figure~\ref{fig:all} shows the box plot distribution of metrics for Gemini-1.5 across these five groups, we can find that \textbf{the group with lower visual saliency has lower SSIM and position similarity (Limitation 3).}  Although the clip and text similarity fluctuates among different groups, as shown in Figure~\ref{fig:box1}, Figure~\ref{fig:box2} shows that the SSIM and position similarity significantly increases as the visual saliency increases. As shown in Figure~\ref{fig:box2}, the group [0.2, 1) shows the highest metrics, while the group [0, 0.025) shows the lowest metrics. This demonstrates that MLLMs are more likely to capture structural and positional features for samples with high visual saliency.


\begin{tcolorbox}[colback=gray!20, colframe=gray!20, width=\columnwidth]
\textbf{Improvement 3: Visual Saliency Enhancement (VSE).} By cropping the image to increase the proportion of the interactive part, VSE makes the model to better perceive the interaction area.
\end{tcolorbox}

\begin{figure}[ht]
    \centering
    \includegraphics[width = .32\textwidth]{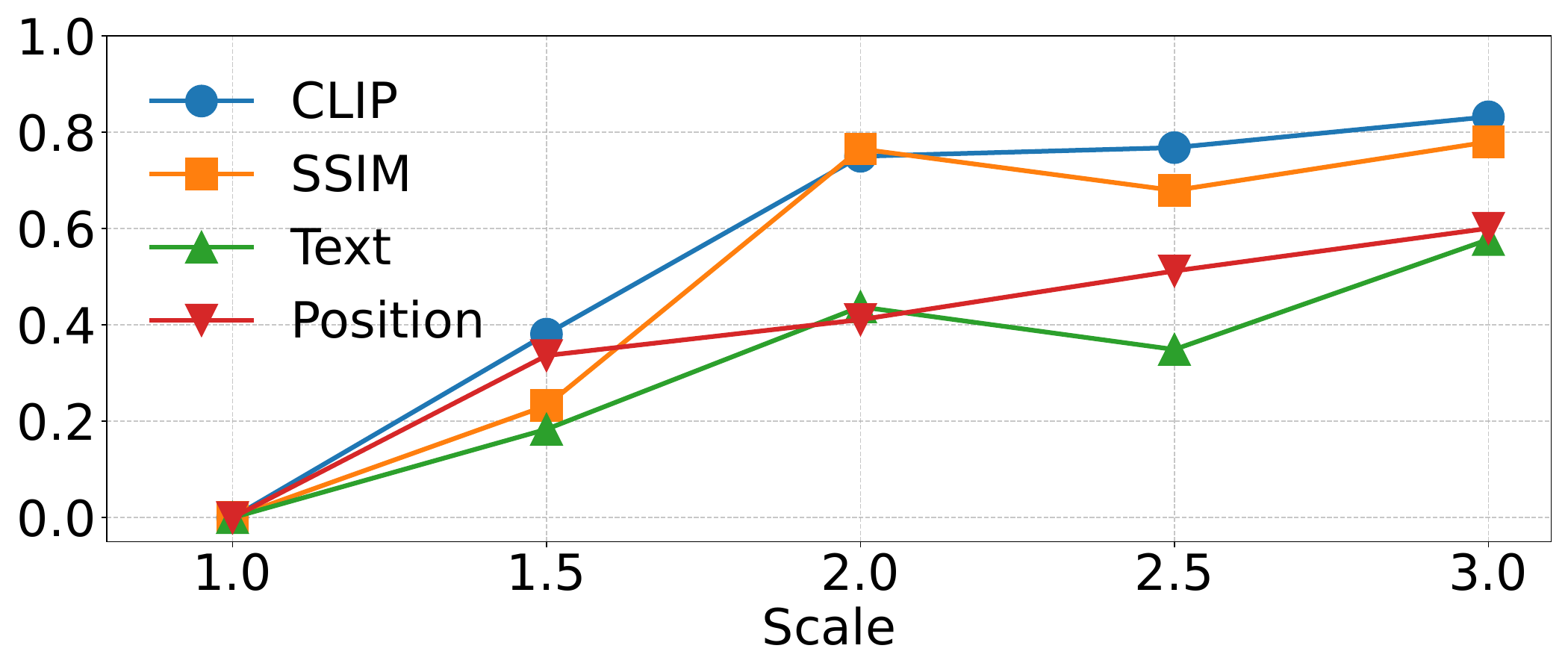}
    \caption{Metrics under different magnification.}
    \label{fig:scale}
\end{figure}

We then randomly sample 10 webpages from failure cases and crop the screenshots to increase the visual saliency of the interactions in the webpages (for example, if the webpage is cropped to $\frac{1}{2}$ of the original, the visual saliency of the interaction will be doubled). Figure~\ref{fig:scale} shows the relationship between the magnification factor and the metrics of generation results. We observe that: when the magnification factor is set to 1, all evaluation metrics yield values of 0, indicating the unsuccessful interaction generation. Upon increasing VS by 1.2 times, the model is able to reproduce interactions, but with relatively low metric scores. As the magnification factor increases from 1.2 to 3, we observe substantial improvements in performance metrics: the CLIP and SSIM similarities approach 0.8, while text and position similarities reach approximately 0.6. This suggests that models overcome the original failures.

Although increasing magnification may result in some UI areas being cut off, this can be effectively addressed through a two-step generation approach:(1) Step 1: Generate the interactive components within the magnified, focused area. (2) Step 2: Generate the remaining cut-off UI portions separately. (3) Step 3: Combine both parts to produce the complete webpage.

Using the above method, we sampled 10 webpages with magnification factors ranging from 1x to 3x and observe the performance. Table~\ref{tab:mag_full} shows that full-page similarity scores remain consistent. So we believe \textbf{the models' performance on the full UI page remains stable with the increased magnification factor.}


\begin{table}[htbp]
\centering
\caption{The full page performance under different magnification}
\begin{tabular}{l|ccccc}
\toprule
Magnification & 1 & 1.5 & 2 & 2.5 & 3 \\
\midrule
CLIP & 0.826 & 0.842 & 0.838 & 0.843 & 0.836 \\
SSIM & 0.652 & 0.657 & 0.643 & 0.654 & 0.653 \\
Text & 0.657 & 0.659 & 0.649 & 0.665 & 0.659 \\
\bottomrule
\end{tabular}
\label{tab:mag_full}
\end{table}




\subsection{The Impact of Different Modalities}

\begin{table}[ht]
\centering
\small
\caption{Performance of GPT-4o with different modality inputs. \textbf{Bold values} are the best performance and \underline{underlined values} are the second-best performance.}
\label{tab:des}
\begin{tabular}{@{}c|c|cccc@{}}
\toprule
Prompt & Modality  & CLIP            & SSIM            & Text            & Position        \\ \cmidrule(l){1-6}
\multirow{3}{*}{Direct} & V                       & 0.3737          & 0.1793          & \underline{0.2539}          & 0.3951          \\
                        & T                       & \underline{0.4174}          & \underline{0.4067}          & 0.2316          & \underline{0.4293}          \\
                        & V+T                     & \textbf{0.6735}          & \textbf{0.5612} & \textbf{0.3919}          & \textbf{0.7157}          \\ \cmidrule(l){1-6} 
\multirow{3}{*}{CoT}    & V                       & 0.3871          & \underline{0.3101}          & 0.2433          & 0.4461          \\
                        & T                       & \underline{0.5579}          & 0.1828          & \underline{0.3045}          & \underline{0.5465}          \\
                        & V+T                     & \textbf{0.6440}           & \textbf{0.4800}            & \textbf{0.4287} & \textbf{0.7080}           \\ \cmidrule(l){1-6} 
\multirow{3}{*}{Mark}   & V                       & \underline{0.5015}          & \textbf{0.4520}           & \underline{0.3389}          & \underline{0.5025}          \\
                        & T                       & 0.4613          & \underline{0.4454}          & 0.2805          & 0.4810           \\
                        & V+T                     & \textbf{0.6923} & 0.4336          & \textbf{0.4248}          & \textbf{0.7469} \\ \bottomrule
\end{tabular}
\end{table}


MLLMs' UI code generation effectiveness hinges on interaction comprehension, with complex or visually subtle interactions being particularly challenging when using images alone. Natural language descriptions can complement visual inputs. To investigate the impact of different input signals, we conduct experiments on GPT-4o using 10 randomly selected webpages from failure cases. Human annotators provide textual descriptions for each interaction (e.g., "clicking the login button triggers a new window with two input boxes"). We evaluate three settings: visual input only (V), textual description only (T), and combined visual-textual input (V+T). Table~\ref{tab:des} shows that \textbf{visual-only (V) and text-only (T) inputs exhibits unsatisfactory performance (Limitation 4)}, the combined approach (V+T) consistently outperforms single-modality inputs across all prompt types, indicating complementary benefits.

\begin{tcolorbox}[colback=gray!20, colframe=gray!20, width=\columnwidth]
\textbf{Improvement 4: Visual and Textual Description Combination.} Combined visual and textual inputs can optimize MLLMs' \task \ performance.
\end{tcolorbox}

\subsection{Ablation Study}
To investigate whether the improvement methods address complementary weaknesses or have overlapping benefits, we conducte the ablation experiments on GPT4o, which evaluates the combinations of all proposed enhancement strategies. The results in Table~\ref{tab:ablation} show that combining Interactive Element Highlighting (IEH), Failure-aware Prompt (FAP), Visual Saliency Enhancement (VSE) and Textual Description (TD) can achieve the best performance, and removing any one of them leads to a decrease in performance (\ding{55} denotes "without the module", \checkmark indicates "with the module".)

\begin{table}[ht]
    \centering
    \caption{Combinations of proposed improvements and correspoding performance.}
    \label{tab:ablation}
    \begin{tabular}{cccccccc}
        \toprule
        IEH & FAP & VSE & TD & CLIP & SSIM & Text & Position \\
        \midrule
        \ding{55} & \ding{55} & \ding{55} & \ding{55} & 0.5700 & 0.4891 & 0.3652 & 0.5802 \\
        \checkmark & \ding{55} & \ding{55} & \ding{55} & 0.5968 & 0.5456 & 0.4508 & 0.6408 \\
        \ding{55} & \checkmark & \ding{55} & \ding{55} & 0.6072 & 0.5405 & 0.4580 & 0.6452 \\
        \ding{55} & \ding{55} & \checkmark & \ding{55} & 0.6328 & 0.5723 & 0.4629 & 0.6874 \\
        \ding{55} & \ding{55} & \ding{55} & \checkmark & 0.6547 & 0.5681 & 0.4732 & 0.7024 \\
        \ding{55} & \checkmark & \checkmark & \checkmark & 0.6584 & 0.5938 & 0.4852 & 0.7027 \\
        \checkmark & \ding{55} & \checkmark & \checkmark & 0.6627 & 0.5987 & 0.4764 & 0.7128 \\
        \checkmark & \checkmark & \ding{55} & \checkmark & 0.6737 & 0.6136 & 0.4828 & 0.7269 \\
        \checkmark & \checkmark & \checkmark & \ding{55} & 0.6532 & 0.5823 & 0.4621 & 0.6987 \\
        \checkmark & \checkmark & \checkmark & \checkmark & \textbf{0.6937} & \textbf{0.6251} & \textbf{0.5187} & \textbf{0.7371} \\
        \bottomrule
    \end{tabular}

\end{table}

\subsection{Human Evaluation of Interaction2Code Tool}
To demonstrate the practical use of the Interaction2Code paradigm, we conduct a user study to asses the impact of our task on dynamic website developers.

\textbf{Participant and preparation.} We hired four PhD students with similar front-end development experience to complete four interaction tasks within 30 mins. Of these participants, two utilized our Interaction2Code tool (described in Appendix H), while the other two used LLMs but did not access the Interaction2Code tool.

\textbf{Study setting.} We recorded the time taken to complete tasks. We also invite other two front-end development experts evaluate their implementations on a scale of 1 to 5, with higher scores indicating better implementations.

\textbf{Result.} Table~\ref{tab:interaction2code_comparison} presents the time costs and performance metrics. Our results demonstrate substantial improvements in both development efficiency (over 32.4\%) and implementation quality (over 16.1\%) when using the Interaction2Code Tool. Our user studies demonstrate that Interaction2Code task provides substantial assistance for dynamic webpage developers.

\textbf{Speed-up Mechanism.} Interaction2Code accelerates development by shifting the workflow from "generate-from-scratch" to "review-and-refine": (1) Ready-to-use draft: the tool emits a fully functional webpage that already contains the correct HTML structure, CSS layout, and JavaScript event handlers for the specified interaction. The developer do not need to write code from scratch. (2) Reduced lookup time: developers no longer spend minutes searching docs for the api usage like ``onClick'' signature, event attributes, etc. (3) Faster iteration loop: developers skip writing many lines of codes and jump straight to testing the generated interaction: either approve it or apply minor adjustments.

\begin{table}[ht]
    \caption{Task Time and Similarity With and Without Interaction2Code. w. denotes results with Interaction2Code tool, w/o denotes results without the tool.}
    \label{tab:interaction2code_comparison}
    \centering
    \begin{tabular}{lcccc}
        \toprule
        Task & Time (w.) & Time (w/o) & Similarity (w.) & Similarity (w/o) \\
        \midrule
        Task 1 & 323s & 487s & 5   & 4     \\
        Task 2 & 182s & 319s & 4   & 4     \\
        Task 3 & 125s & 189s & 5   & 4     \\
        Task 4 & 486s & 657s & 4   & 3.5   \\
        \midrule
        \textbf{Average} & \textbf{279s} & \textbf{413s} & \textbf{4.5} & \textbf{3.875} \\
        \bottomrule
    \end{tabular}
\end{table}

\section{Threats to Validity}

\textit{Limited context length.} As webpages become more complex with numerous interactions, the input context expands, potentially exceeding the context window constraints of MLLMs (e.g., 128K tokens for GPT-4o). Nevertheless, this limitation can be mitigated by employing iterative generation, progressively producing interactions for a webpage over multiple rounds.

\textit{Model selection.} This study utilizes three prominent Multimodal Large Language Models (MLLMs) to conduct experiments. There are some MLLMs such as Pixtral \cite{agrawal2024pixtral} we don't test, we will test the performance of these models on \task \ task in the future work.

\textit{Unable to handle interactions that require back-end.} Some complex functional interactions (e.g., login, search, etc.) are implemented by server-side scripting languages like Python. The benchmark we collect does not include back-end code; we cannot verify the generation effect of such interactions, but we believe our work is an important step toward generating interactive websites.

\section{Conclusion}

We present the first systematic study of MLLMs' capabilities in generating interactive webpages. We formulate the \textbf{\task} task and establish the \textbf{\benchmark} benchmark. Through comprehensive experiments, we identify four critical limitations: (1) inadequate generation of interaction compared with full page, (2) susceptibility to ten types of failures, (3) poor performance on visually subtle interactions, and (4) insufficient comprehension when limited to single-modality visual descriptions. To address these limitations, we propose four enhancement strategies: interactive element highlighting, failure-aware prompting (FAP), visual saliency enhancement, and the integration of visual-textual descriptions. 

\section{Acknowledgment}

The work described in this paper was supported by two grants from the Research Grants Council of the Hong Kong Special Administrative Region, China: (1) No. CUHK 14209124 of the General Research Fund, and (2) No. SRFS2425-4S03 of the Senior Research Fellow Scheme.
The work was also supported by the Singapore Ministry of Education (MOE) Academic Research Fund (AcRF) Tier 1 grant and the Jingyu Xiao's Hong Kong PhD Fellowship Scheme (No. PF23-87650) under the Research Grants Council of Hong Kong.

\bibliographystyle{IEEEtran} 
\bibliography{ref}

\end{document}